\documentclass[12pt]{iopart}
  \expandafter\let\csname equation*\endcsname\relax
  \expandafter\let\csname endequation*\endcsname\relax
\usepackage{amsmath}
\usepackage{graphicx,eucal}
\usepackage{graphicx}
\usepackage{float}
\usepackage{cite}
\usepackage{subfigure}
\usepackage{color}

\newcommand{\gl}{\mathrel{\raise0.6ex\hbox{$>$\kern-.75em\lower1ex\hbox{$<$}}}}
\newcommand{\agt}{\mathrel{\raise0.6ex\hbox{$>$\kern-.75em\lower1.2ex\hbox{$\sim$}}}}

\begin{document}
\title{Highly Dispersed Networks Generated by Enhanced Redirection}
\author{Alan Gabel$^1$, P~L~Krapivsky$^1$, and S~Redner$^{1,2}$}
\address{$^1$Department of Physics, Boston University, Boston, Massachusetts 02215, USA}
\address{$^2$Santa Fe Institute, 1399 Hyde Park Road, Santa Fe,
  New Mexico 87501, USA}

\begin{abstract}
  We analyze growing networks that are built by enhanced redirection.  Nodes
  are sequentially added and each incoming node attaches to a randomly chosen
  `target' node with probability $1-r$, or to the parent of the target node
  with probability $r$.  When the redirection probability $r$ is an
  increasing function of the degree of the parent node, with $r\to 1$ as the
  parent degree diverges, networks grown via this \emph{enhanced} redirection
  mechanism exhibit unusual properties, including: (i) multiple macrohubs,
  i.e., nodes with degrees proportional to the number of network nodes $N$;
  (ii) non-extensivity of the degree distribution in which the number of
  nodes of degree $k$, $N_k$, scales as $N^{\nu-1}/k^{\nu}$, with $1<\nu<2$;
  (iii) lack of self-averaging, with large fluctuations between individual
  network realizations.  These features are robust and continue to hold when
  the incoming node has out-degree greater than 1 so that networks contain
  closed loops.  The latter networks are strongly clustered; for the specific
  case of double attachment, the average local clustering coefficient is
  $\langle C_i \rangle =4\ln2-2=0.77258\dots$.
\end{abstract}

\maketitle

\section{Introduction}

Models for the growth of complex networks often involve mechanisms that are
based on \emph{global} knowledge of the network.  For example, in
preferential attachment~\cite{Y25,S55,BA99,KRL00,DMS00,DM03,N10}, nodes are
added sequentially and each links to existing target nodes in the network
according to an attachment rate $A_k$ that is an increasing function of the
degree $k$ of the target node.  According to this rule, incoming nodes must
`know' the degree distribution of the entire network to correctly choose a
target node.  In real networks, however, it is not feasible that any new node
has such detailed global knowledge.

The impracticality of implementing a growth rule based on global knowledge
has motivated alternatives to preferential attachment that rely on the
incoming nodes exploiting only \emph{local} knowledge of a small portion of
the network.  Examples include attachment via spatial
locality~\cite{FKP02,CBMR04,B11} and node similarity~\cite{PKSBK12}.  In this
work, we focus on the local growth rule that exploits
\emph{redirection}~\cite{kum,KR01,V03,RA04,KR05,LA07,BK10}.  Here, each
incoming node selects a target node at random and links either to this target
node (probability $1-r$), or  to parent of the target node (probability
$r$).  This redirection rule is based on the network being directed so that
the parent(s) of any node is well defined.  If each new node has only one
outgoing link, redirection produces networks with a tree topology; it is
straightforward to extend redirection to allow each incoming node to attach
to more than one node in the network~\cite{RA04}.

The surprising feature of redirection with a fixed redirection probability
$r$ is that it is mathematically equivalent to the global growth rule of
shifted linear preferential attachment, where the rate of attaching a new
node to a pre-existing node of degree $k$ is $A_k=k+\lambda$ with
$\lambda=r^{-1}-2$, see \cite{KR01}.  Redirection is also highly efficient
because one only needs to select a random node and identify its parent to add
a node to the network.  The time to create a network of $N$ nodes via
redirection therefore scales linearly with $N$.

\begin{figure}[ht]
\begin{center}
\includegraphics[width=0.4\textwidth]{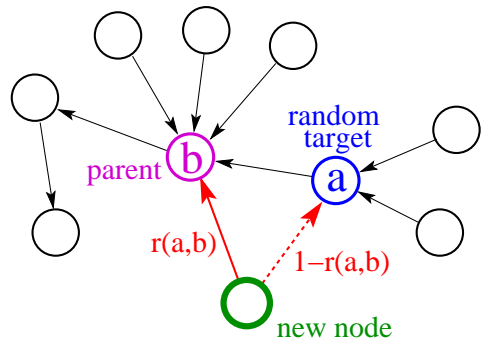}
\caption{Illustration of enhanced redirection.  The redirection probability
  $r(a,b)$ depends on the degree $a$ of a randomly chosen target node and the
  degree $b$ of its parent, with $r(a,b)$ an increasing function of $b$ and
  $r\to 1$ as $b\to\infty$.  In this example, the degree of the parent node
  is large, so the new node is likely to attach to the parent.  }
  \label{rules}
\end{center}
\end{figure}

The utility of redirection as a simple and efficient procedure that is
equivalent to linear preferential attachment motivates us to exploit models
that use slightly more comprehensive (but still local) degree information
around the target node.  Specifically we allow the redirection probability
$r(a,b)$ to depend on the degrees of the target and parent nodes, $a$ and $b$
respectively (Fig.~\ref{rules}).  In \emph{hindered redirection}, $r(a,b)$ is
a decreasing function of the parent degree $b$, a rule that leads to
sub-linear preferential attachment growth~\cite{GR13}.  In this work, we
investigate the complementary situation of \emph{enhanced redirection}, for
which the redirection probability $r$ is an increasing function of the parent
degree $b$, with $r\to 1$ as $b\to\infty$.  This seemingly-innocuous
redirection rule gives rise to networks with several intriguing and
practically relevant properties:

\begin{enumerate}

\item {\bf \emph{Appearance of multiple macrohubs}:} Macrohubs are nodes
  whose degrees are a finite fraction of $N$.  While macrohubs arise in other
  models~\cite{KRL00,KR01,KK08,CS03,BB01}, the resulting networks are
  singular, with nearly all nodes attached to a single macrohub.  In the
  cases of superlinear preferential attachment, where $A_k\sim k^\gamma$ with
  $\gamma>1$~\cite{KRL00,KR01,KK08}, and in the fitness model, where the
  attachment rate is proportional to both the degree $k$ and fitness of the
  target~\cite{BB01,KR02}, a single macrohub arises that is connected to
  almost all other nodes of the network.  In contrast, enhanced redirection
  networks are highly disperse (Fig.~\ref{typical}), with interconnected
  hub-and-spoke structures that are reminiscent of airline route
  networks~\cite{N10,CS03,BO99,H04,GMTA05}. 

\begin{figure}[ht]
\begin{center}
\medskip
\subfigure[]{\includegraphics[width=0.31\textwidth]{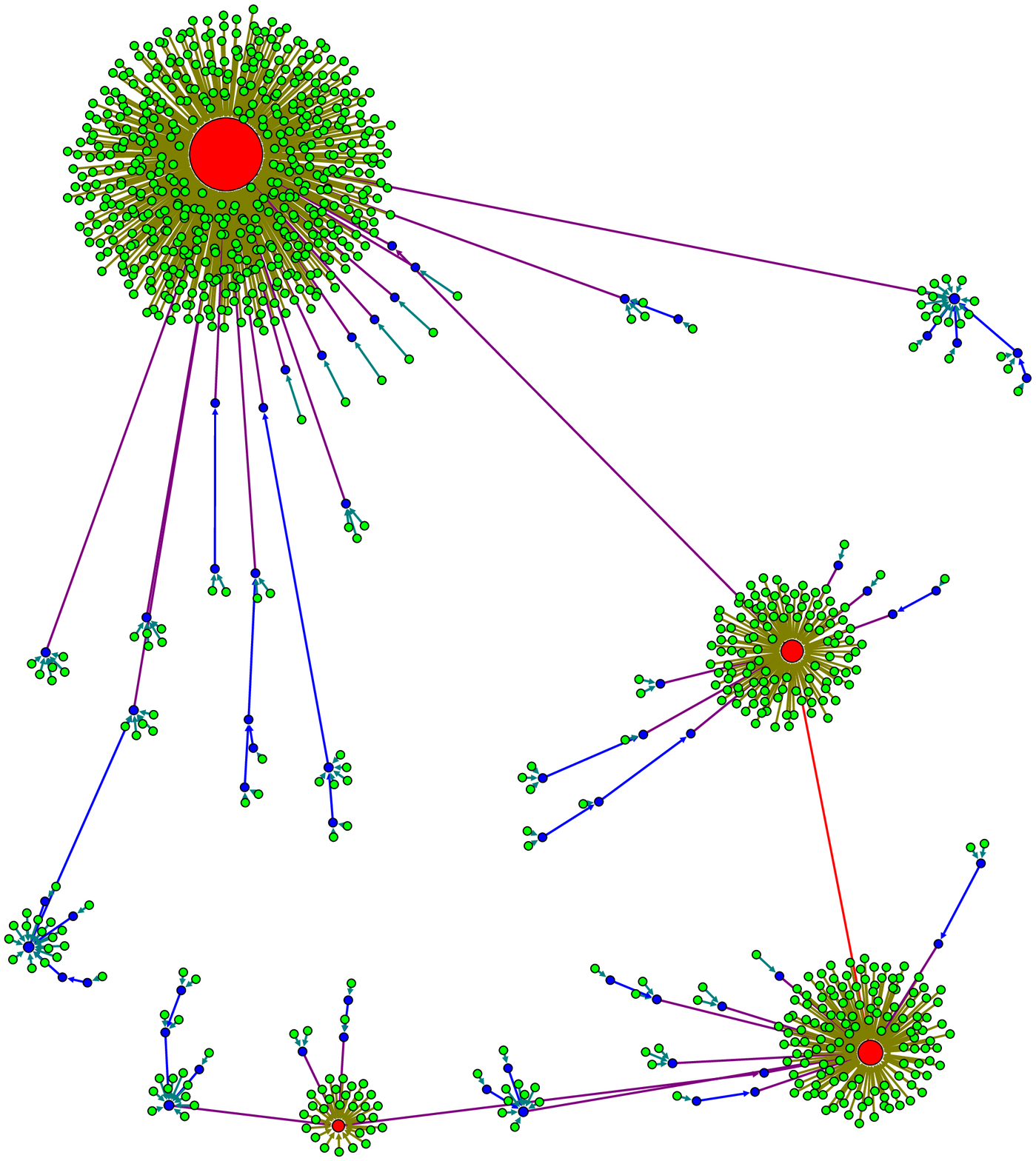}} \quad
\subfigure[]{\includegraphics[width=0.31\textwidth]{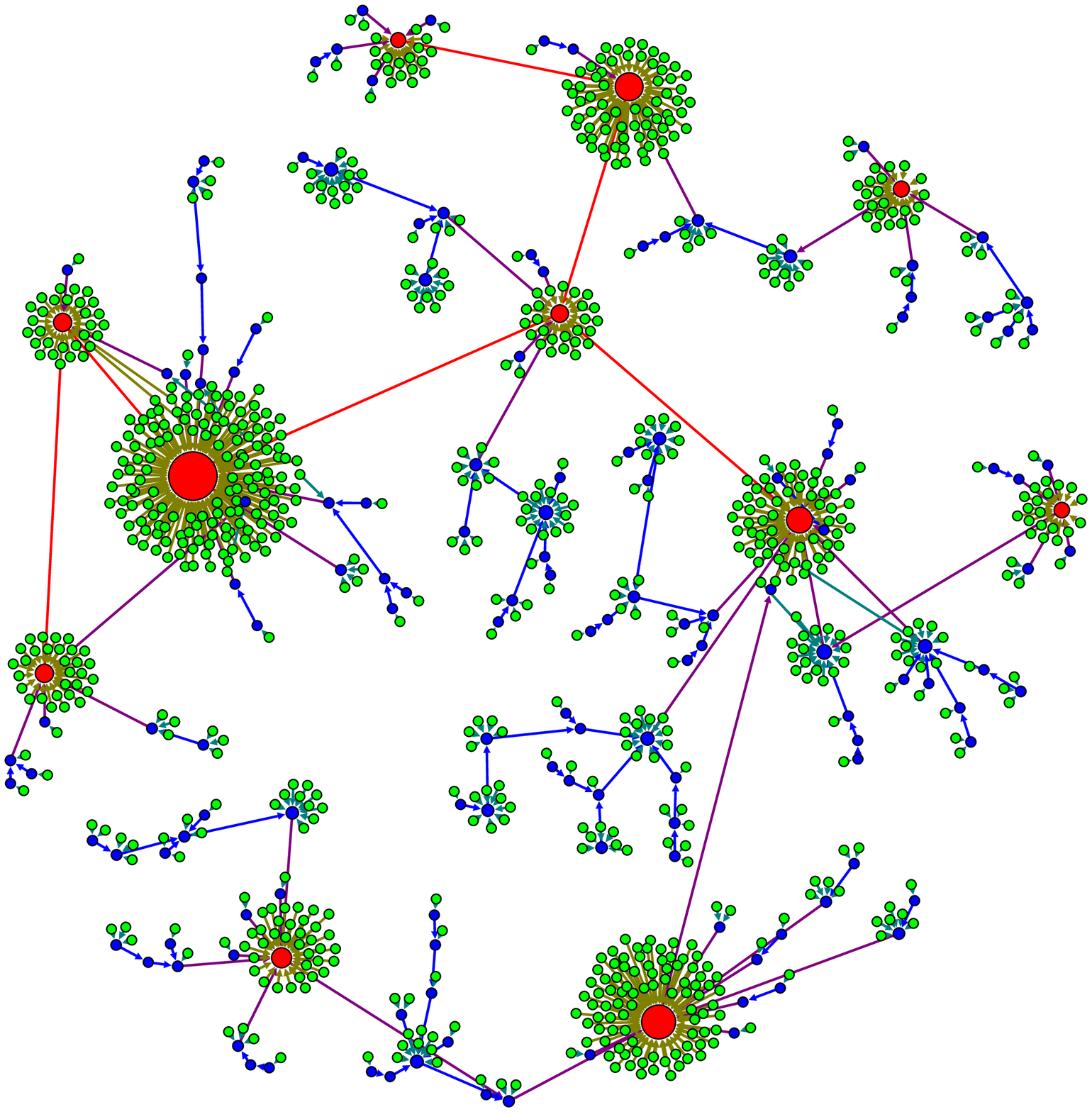}} \quad
\subfigure[]{\includegraphics[width=0.31\textwidth]{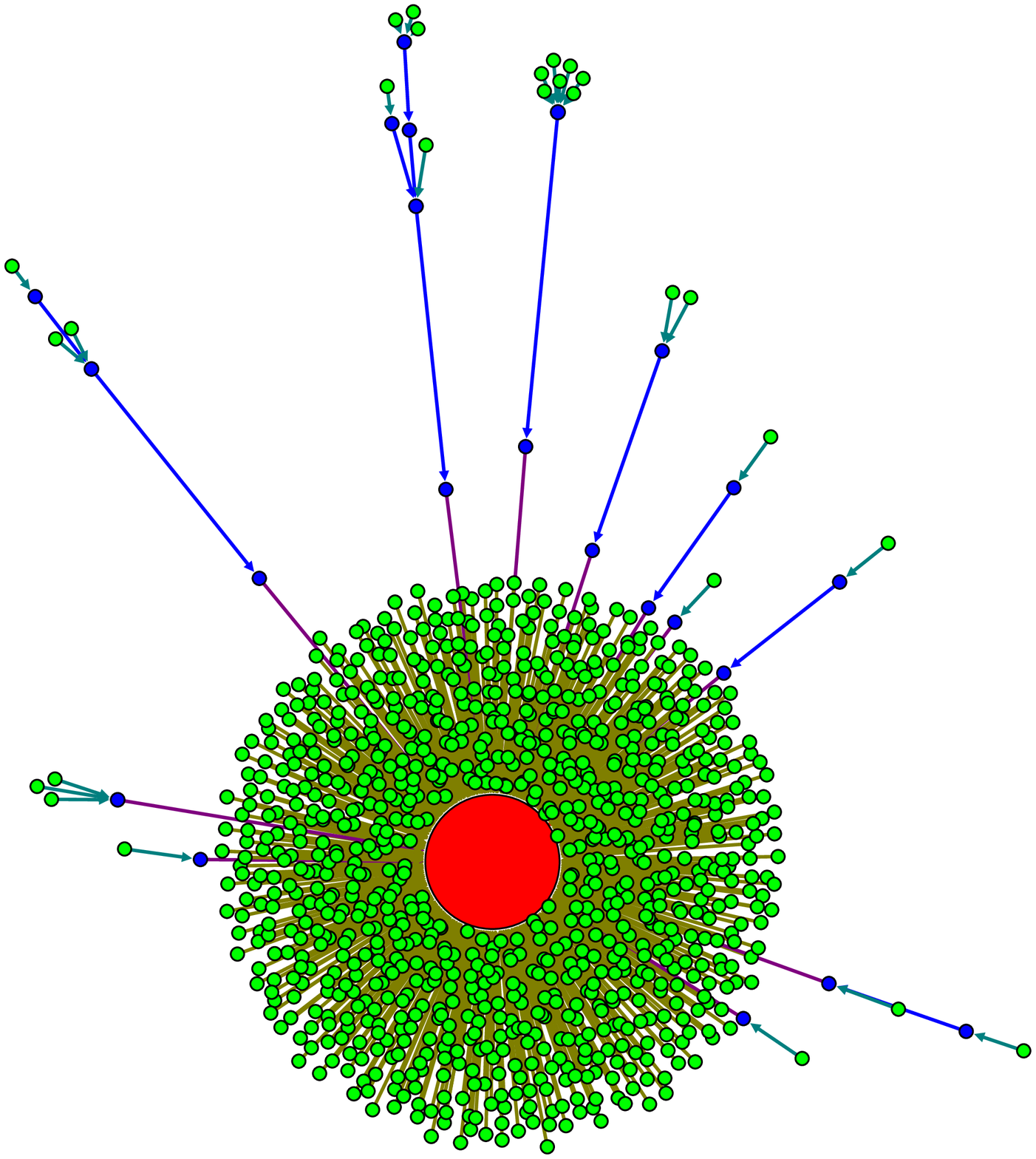}}
\caption{Enhanced redirection networks of $N=10^3$ nodes for
  $\lambda=\frac{3}{4}$ (see Eq.~\eqref{r}) starting from the same initial
  state.  (a) Maximum degree $k_{\rm max}=548$, $\mathcal{C}=66$ core ($k\geq
  2$) nodes, and maximum depth $D_{\rm max}=10$.  (b) $k_{\rm
    max}=\mathcal{C}=154$, $D_{\rm max}=12$ (smallest $k_{\rm max}$ out of
  $10^3$ realizations).  (c) $k_{\rm max}=963$, with $\mathcal{C}=23$ and
  $D_{\rm max}=6$ (largest $k_{\rm max}$ out of $10^3$ realizations).  Green:
  nodes of degree 1, blue: degrees 2--20, red: degree $>20$.  The link color
  is the average of the endpoint node colors.}
  \label{typical}
\end{center}
\end{figure}

\item {\bf \emph{Non-extensivity}:} In many sparse networks, the degree
  distribution is extensive, with the number of nodes of degree $k$, $N_k$,
  proportional to $N$.  This happens, for example, in linear preferential
  attachment where additionally the degree distribution has an algebraic
  tail, $N_k\sim N/k^{\nu}$ for $k\gg 1$, with $\nu>2$.  In contrast,
  enhanced redirection leads to the non-extensive scaling
\begin{equation}
\label{NkN}
 N_k\sim \frac{N^{\nu-1}}{k^{\nu}} \qquad \text{with}\quad \nu<2.
\end{equation}
The allowed range of the exponent $\nu$ is key.  While past empirical studies
have observed networks with degree exponent in the range
$1<\nu<2$~\cite{KBM13} (see Table~1 for some examples), the range $1<\nu<2$
is mathematically inconsistent for sparse networks because it leads to a
divergent average degree as $N\to\infty$ whenever the degree distribution
obeys the standard scaling $N_k\sim Nk^{-\nu}$.  A simple resolution of this
dilemma is to relax the hypothesis of extensivity.  We shall see that in
enhanced redirection almost all nodes have degree 1 (leaves).  More
precisely, the number $\mathcal{C}\equiv N-N_1$ of core nodes (nodes with
degree $>1$) grows \emph{sub-linearly} with $N$, namely as $N^{\nu-1}$ with
the exponent $\nu$ in the range $1<\nu<2$.  All $N_k$ with $k\geq 2$ also
grow as $N^{\nu-1}$.  This anomalous scaling can therefore be summarized as
follows:
\begin{equation}
\label{scaling-gen}
\mathcal{C} = N\!-\!N_1 \simeq c_1 N^{\nu-1}\,,\qquad 
N_k\simeq c_k N^{\nu-1} \quad \text{for} \quad k\geq 2 
\end{equation}
where $c_k$ are constants.  This scaling satisfies the sum
rule $\sum_{1\leq k\leq N} N_k=N$ and leads to a finite average degree
$\langle k \rangle$ without imposing an artificial cutoff in the degree
distribution.

\begin{table}[ht]
\begin{center}
\label{compare}
\begin{tabular}{|c|c|c|c|}
\hline
{\bf Network}                 & exponent $~\nu$  & average degree & network size \\
\hline
Orkut                                     &  1.27       & 76.281                 & 3,072,441\\
\hline
Catster Friendships                &  1.36             & 72.803                 & 149,700\\
\hline
Dogster Friendships               &  1.40             & 40.048                 & 426,820\\
\hline
arXiv hep-ph                      &  1.47               & 224.14                & 28,093\\
\hline
arXiv hep-th                       &  1.47               & 213.44                 & 22,908\\
\hline
Wikipedia conflict                & 1.50                & 34.644                 & 118,100\\
\hline
Hamsterster full                     &  1.52             & 13.711                 & 16,630\\
\hline
Hamsterster Friendships     &  1.54             & 13.491                 & 1,858\\
\hline
Flickr                         &  1.73               & 43.742                & 105,938\\
\hline
Internet topology               &  1.86               & 9.8618                 & 34,761\\
\hline
\hline
\hline
\hline
Wikipedia, Italian            & 1.48                 & 28.457                  & 1,204,009\\
\hline
Wikipedia, German        & 1.50               & 28.811                 & 2,166,669\\
\hline
LiveJournal                           & 1.56                & 28.465                 & 4,847,571\\
\hline
Wikipedia, French           & 1.62                & 22.165                 & 2,212,682\\
\hline
OpenFlights                     & 1.79                & 20.756                 & 2,939\\
\hline
\end{tabular}
\caption{\small Networks with degree exponent $\nu<2$.  All examples are
  simple graphs (at most one link between any node pair); the first 10 are
  undirected and the remainder are directed.  Data at
  http://konect.uni-koblenz.de/networks.  }
\end{center}
\end{table}

\item {\bf \emph{Lack of self-averaging}:} Different realizations of enhanced
  redirection are visually diverse when starting from the same initial
  condition (Fig.~\ref{typical}).  Basic network measures, such as the number
  of nodes of fixed degree, $N_k$ with any $k\geq 2$, or the number of core
  nodes $\mathcal{C}$, vary significantly between realizations and do not
  converge as $N\to\infty$.  For instance, the ratio of the mean deviation to
  the average, $\sqrt{\langle \mathcal{C}^2\rangle-\langle
    \mathcal{C}\rangle^2}/\langle \mathcal{C}\rangle$, converges to a
  positive constant when $N\to\infty$, thereby manifesting the lack of
  self-averaging.  In contrast, preferential attachment networks do
  self-average, as the relative deviations in $N_k$ or $\mathcal{C}$
  systematically decrease as $N$ increases~\cite{KR02f}.

\end{enumerate}

In the next section we formally define our enhanced redirection models.  In
Sect.~\ref{DD} we provide analytical and numerical arguments that justify the
properties (i), (ii), (iii) given above.  Some of these arguments
substantially extend our findings that were reported in Ref.~\cite{GKR_13}.
Most results in Sect.~\ref{SS}, and all results in Sect.~\ref{DA} about
enhanced redirection with multiple attachments, are new.

\section{Enhanced Redirection Model}
\label{model:def}

We define the initial network to be a single node that is linked to itself,
so that the root node is its own parent and its own child.  The initial
conditions have a weak and mostly quantitative influence on asymptotic
network properties.  Thus we shall generally use the above simple initial
condition; we will explicitly define other initial conditions in the few
cases where such a modification is more amenable to analysis.
  
Links are directed so that the parent and children of any node are well defined.  
In Sects.~\ref{DD}--\ref{SS} 
we investigate models in which each node has out-degree equal to 1, and thus
a unique parent.  This growth rule produces tree networks if the starting
network is a tree.  Our networks are trees with the exception of the initial
self loop.

Nodes are introduced one by one.  Each incoming node first picks a random
target node.  If the degrees of the target and parent nodes are $a$ and $b$,
respectively, then the new node (see Fig.~\ref{rules})
\begin{enumerate}
\itemsep -0.3ex
\item[(i)] attaches to the target with probability $1-r(a,b)$;
\item[(ii)] or attaches to the parent of the target with probability $r(a,b)$.
\end{enumerate}

Two natural (but by no means unique) choices for the redirection probability
are
\begin{equation}
\label{r}
r(a,b)=1-b^{-\lambda},\qquad\qquad r(a,b)=\frac{a^\lambda}{a^\lambda+b^\lambda},\qquad \lambda>0\,.
\end{equation}
Our results are robust with respect to the form of the redirection
probability, as long as $r(a,b)\to 1$ as $b\to\infty$.  For concreteness, we
focus on the redirection probability $r(a,b)=1-b^{-\lambda}$.
In~\ref{app:star} we compare some results for this case with corresponding
results for the redirection probability
$r(a,b)=a^\lambda/(a^\lambda+b^\lambda)$.  This comparison indicates that the
two models are qualitatively the same.

\section{Degree Distribution and Lack of Self-Averaging}
\label{DD}

\begin{figure}[ht]
\begin{center}
\subfigure[]{\includegraphics[width=0.45\textwidth]{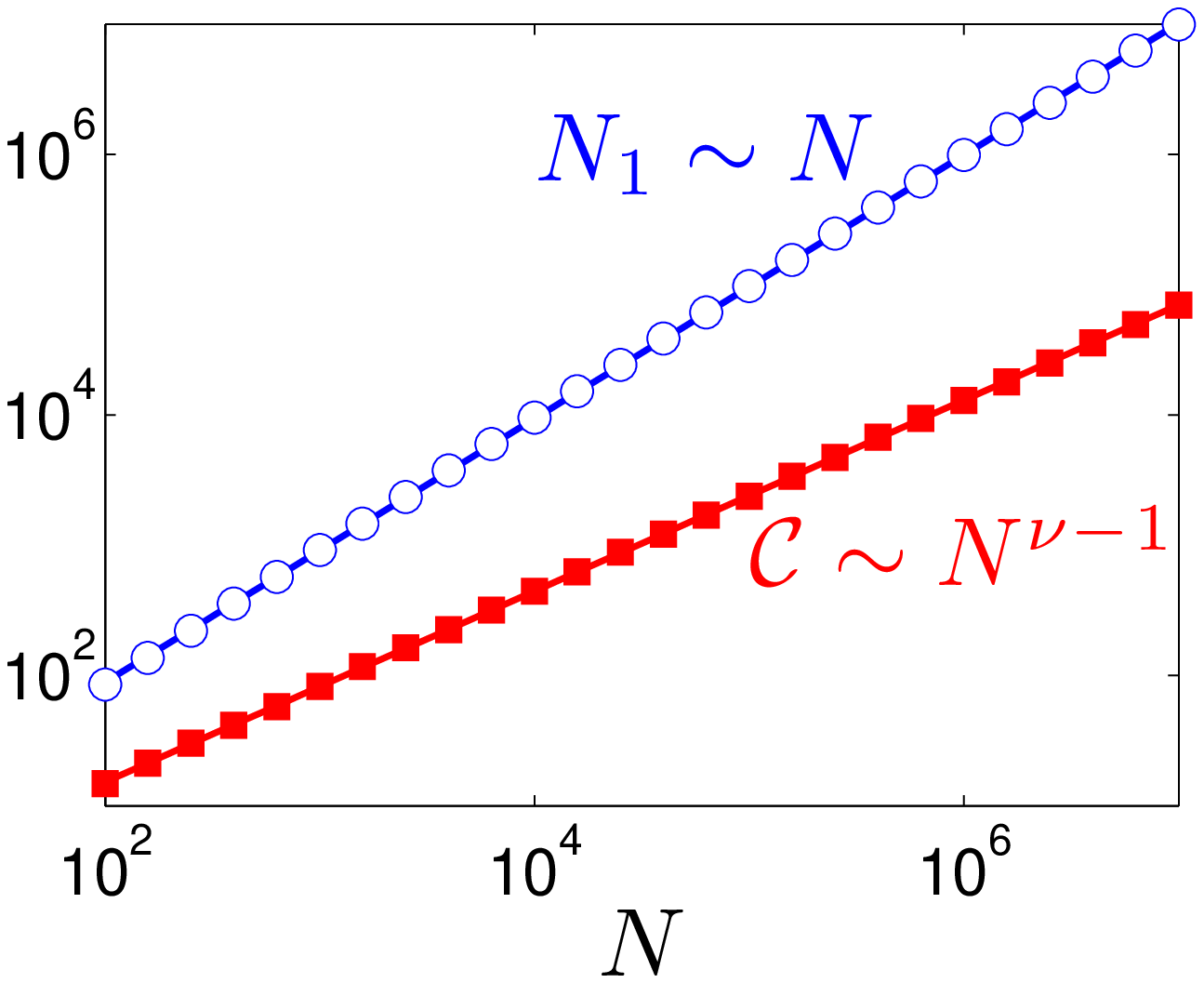}}\qquad
\subfigure[]{\includegraphics[width=0.45\textwidth]{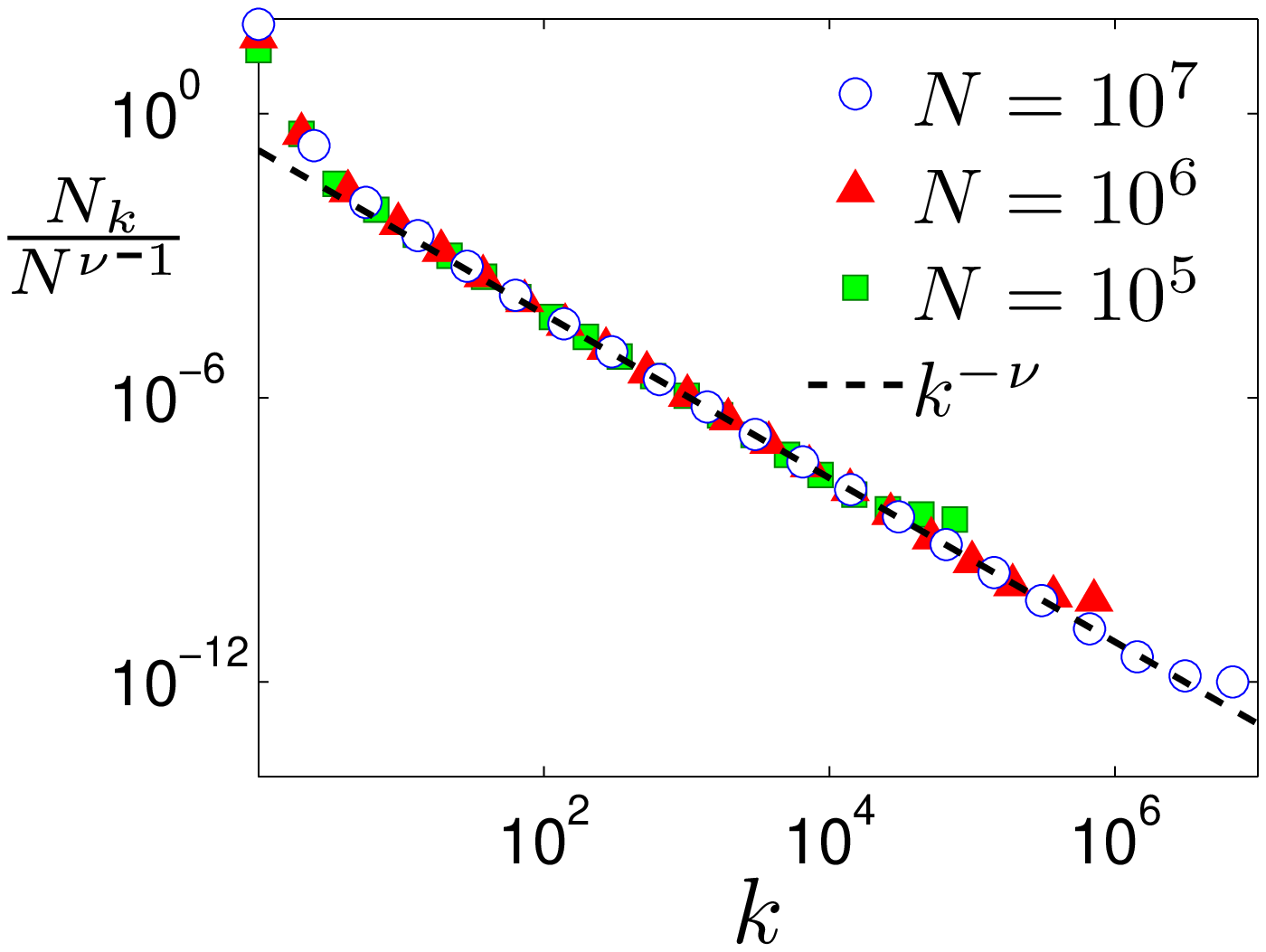}}
\caption{(a) $N_k$ versus $N$ and (b) $N_k/N^{\nu-1}$ versus $k$ for enhanced
  redirection with $\lambda=\frac{3}{4}$ and $\nu=1.73$ (determined
  numerically; see Fig.~\ref{nu}).  Data are based on $10^4$ realizations,
  with equally-spaced bins on a logarithmic scale in (b).  The lines in (a)
  show the prediction of Eq.~\eqref{scaling-gen}, while the line in (b) shows
  the $k$ dependence from the numerical solution of \eqref{ck}.  }
  \label{NkVsN}
\end{center}
\end{figure}

\begin{figure}[ht]
\begin{center}
\includegraphics[width=0.45\textwidth]{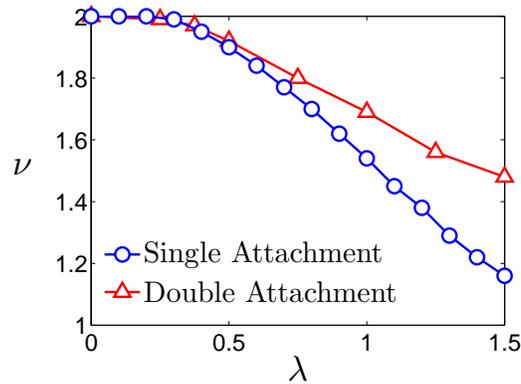}
\caption{Degree distribution exponent $\nu$ versus $\lambda$ for enhanced
  redirection with single attachment (\textcolor{blue}{$\circ$}) and double attachment
  (\textcolor{red}{$\Delta$}).  Each data point is determined from fits of
  $N_k$ versus $N$, as in Figs.~\ref{NkVsN}(a) and \ref{NkDouble}. }
  \label{nu}
\end{center}
\end{figure}

We first study the degree distribution $\langle N_k\rangle$, the average
number of nodes of degree $k$; to avoid notational clutter we drop the angle
brackets henceforth.  Simulation results clearly show that the degree
distribution has the anomalous scaling behaviors given by
Eq.~\eqref{scaling-gen} (Fig.~\ref{NkVsN}).  The exponent $\nu$ depends on
the redirection parameter $\lambda$, but is always less than 2
(Fig.~\ref{nu}) so that the degree distribution decays very slowly in $k$.
Because $\nu<2$, Eq.~\eqref{scaling-gen} implies that the number of nodes of
degree $1$ grow more rapidly with $N$ than the number of core nodes.  Thus,
visually, a typical network is dominated by its leaves.

We employ the master equation approach to understand the anomalous scaling in
Eq.~\eqref{scaling-gen}.  The degree distribution evolves according to
\begin{align}
\label{master1}
\frac{dN_k}{dN}&=\frac{(1\!-\!f_{k-1})N_{k-1}-(1\!-\!f_k)N_k}{N} 
 +\frac{(k\!-\!2)t_{k-1}N_{k-1}-(k\!-\!1)t_kN_k}{N}+\delta_{k,1}\,.
\end{align} 
Here $f_k$ and $t_k$ are defined as the respective probabilities that an
incoming link is redirected \emph{from} a node of degree $k$, and redirected
\emph{to} a node of degree $k$.  The first ratio in Eq.~\eqref{master1}
accounts for instances in which the incoming node attaches directly to the
target node.  Thus the term $(1-f_k)N_k/N$ gives the probability that
the randomly selected target node has degree $k$ and the incoming node is
\emph{not} redirected.  The term is negative because the target node degree
increases from $k$ to $k+1$ which causes $N_k$ to decrease.  Similarly, the
second ratio corresponds to instances in which the incoming node is
redirected to the parent.  Thus the term $(k-1)t_kN_k/N$ gives the
probability that one of the $(k-1)N_k$ children of a degree $k$ node is
targeted and that the incoming node is redirected to the parent.  The term
$\delta_{k,1}$ arises because each newly added node has degree $1$.

The probabilities $f_k$ and $t_k$ are defined by
\begin{equation}
\label{fk}
f_k=\sum_{b\ge 1} \frac{r(k,b)N(k,b)}{N_k}\,,\qquad \qquad 
t_k=\sum_{a\ge 1}\frac{r(a,k)N(a,k)}{(k-1)N_k}\,.
\end{equation}
Here the correlation function $N(a,b)$ is defined as the number of nodes of
degree $a$ that have a parent of degree $b$.  Thus $f_k$ is the probability
of redirecting \emph{from} a node of degree $k$, averaged over all such
target nodes, and $t_k$ is the probability of redirecting \emph{to} a node of
degree $k$, averaged over all the $(k-1)N_k$ children of nodes of degree $k$.
Defining $\alpha_k=(k\!-\!1)t_k+1-f_k$, Eq.~\eqref{master1} can be written in
the canonical form
\begin{align}
\label{master2}
\frac{dN_k}{dN}=\frac{\alpha_{k-1}N_{k-1}-\alpha_kN_k}{N}+\delta_{k,1}\,.
\end{align} 

Substituting Eq.~\eqref{scaling-gen} into the master equations \eqref{master1}
gives the recursions:
\begin{equation}
\begin{split}
\label{recur}
(\nu-1) c_1 N^{\nu-2} &= \alpha_1\big(1-c_1N^{\nu-2}\big)\,\hskip 1.225in k=1\\
(\nu-1) c_2 N^{\nu-2} &= \alpha_1\big(1-c_1N^{\nu-2}\big) -\alpha_2 c_2
N^{\nu-2}\,\qquad k=2\\
c_k&= \frac{\alpha_{k-1}}{\alpha_k+\nu-1}\, c_{k-1}\hskip 1.26in k\geq 3\,.
\end{split}
\end{equation}
We eliminate the common factor in the first two lines to obtain
$c_2=\nu/(\alpha_2+\nu)$, which, combining with the recursion for $k\geq 3$
gives the product solution
\begin{equation}
\label{ck}
c_k=c_1\,\frac{\nu-1}{\alpha_k}\prod_{j=2}^k \left( \frac{\alpha_j}{\alpha_j+\nu-1} \right).
\end{equation}

For an explicit solution, we need the analytic form for $\alpha_k$, which
requires the probabilities $f_k$ and $t_k$.  For redirection probability
$r(a,b)=1-b^{-\lambda}$, the quantities $f_k$ and $t_k$ reduce to
\begin{equation}
\label{fksimple}
f_k\!=\!\sum_{b\ge 1} \frac{(1-b^{-\lambda})N(k,b)}{N_k}\equiv 1 \!-\!\langle
b^{-\lambda}\rangle\,,\quad
t_k\!=\!\sum_{a\ge 1}\frac{(1-k^{-\lambda})N(a,k)}{(k-1)N_k}=1\!-\!k^{-\lambda}\,,
\end{equation}
where we use the sum rule $\sum_{a\ge1}N(a,k)=(k-1)N_k$.  We now combine
Eq.~\eqref{fksimple} with $\alpha_k=(k\!-\!1)t_k+1-f_k$, to give
$\alpha_k=k-k^{1\!-\!\lambda}+k^{-\lambda}-f_k\to k$ in the large-$k$ limit.
Finally, using $\alpha_k \sim k$ in the product solution \eqref{ck} gives the
asymptotic behavior
\begin{align}
\label{largeK}
  c_k\sim c_1\,\frac{\nu\!-\!1}{k}\,\prod_{j=2}^k \left( \frac{j}{j+\nu-1}  \right)
  &\sim k^{-\nu}\,.
\end{align}
Thus the degree distribution exhibits anomalous scaling, $N_k\sim
N^{\nu-1}/k^{\nu}$, with $\nu<2$, as given in \eqref{NkN}.

Numerical simulations show that the exponent $\nu$ is a decreasing function
of $\lambda$ (Fig.~\ref{nu}).  For $\lambda\rightarrow0$, enhanced
redirection becomes equivalent to random attachment, for which the degree
distribution is extensive $N_k\sim N$ and $\nu\to 2$.  As $\lambda$
increases, attachment to a single node becomes progressively more likely and
$\nu\to 0$.  We also checked that the exponent $\nu$ is not affected by
different initial conditions such as an initial loop of different sizes.
However, finer details of the degree distribution, such as the probability
distribution for the maximal degree and the number of core nodes, do depend
on the initial condition.

One of the visually striking features of enhanced redirection networks is
that they display large fluctuations from realization to realization, as are
apparent from the examples in Fig.~\ref{typical}.  To quantify these
fluctuations, let us study $P(N_k)$, the distributions of the number of nodes
of fixed degree $k$.  For networks that are grown by preferential attachment,
this distribution becomes progressively sharper as $N$ increases~\cite{KR02},
as long as the degree is not close to it maximal value.  Thus the
\emph{average} number of nodes of a given degree can be regarded as the set
of variables that fully characterizes the degree distribution.  It is only
the nodes of the highest degree that fail to self average~\cite{KR02_b}.

\begin{figure}[ht]
\begin{center}
\subfigure[]{\raisebox{0 em}{\includegraphics[width=0.45\textwidth]{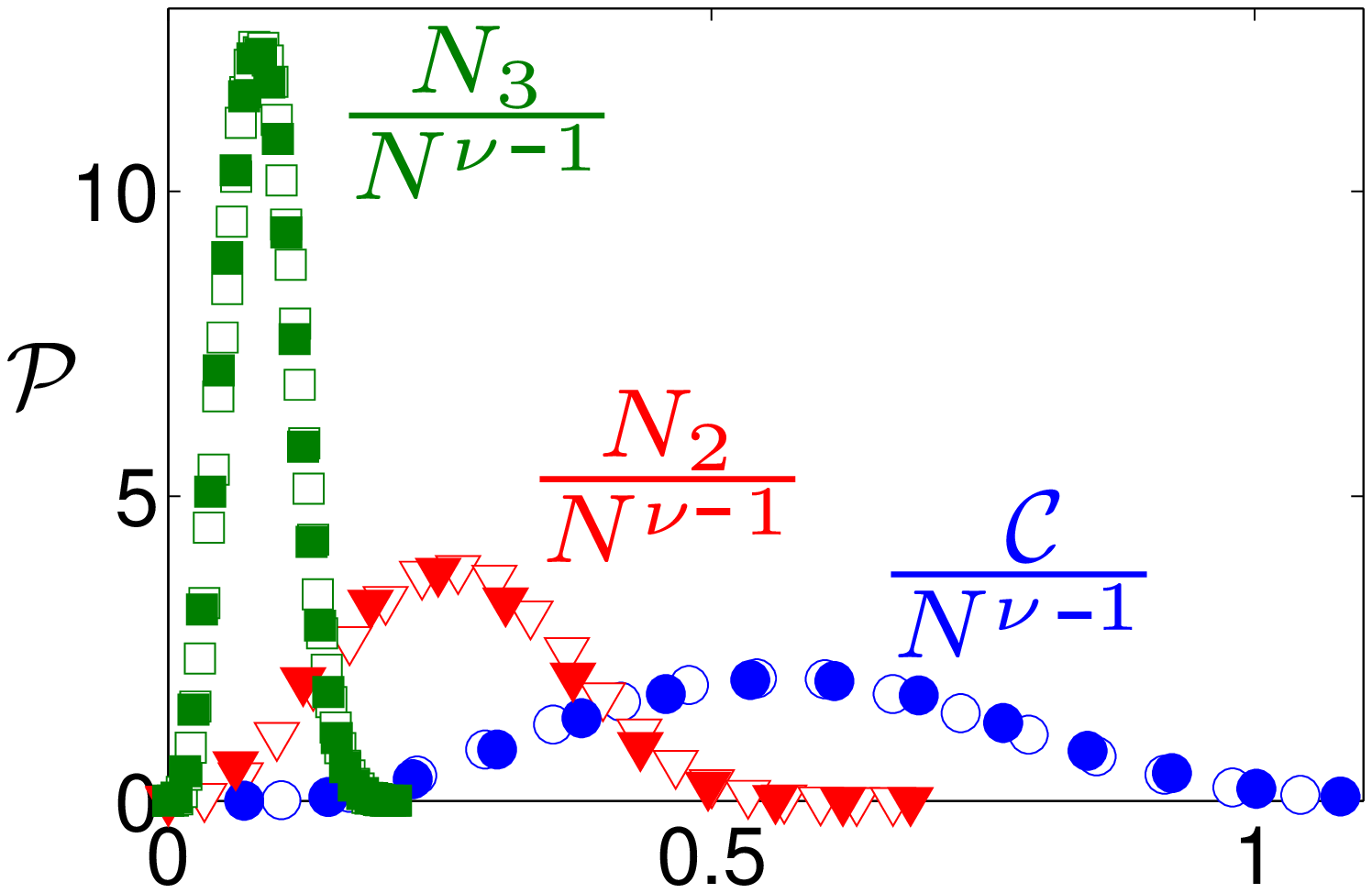}}}\qquad
\subfigure[]{\includegraphics[width=0.45\textwidth]{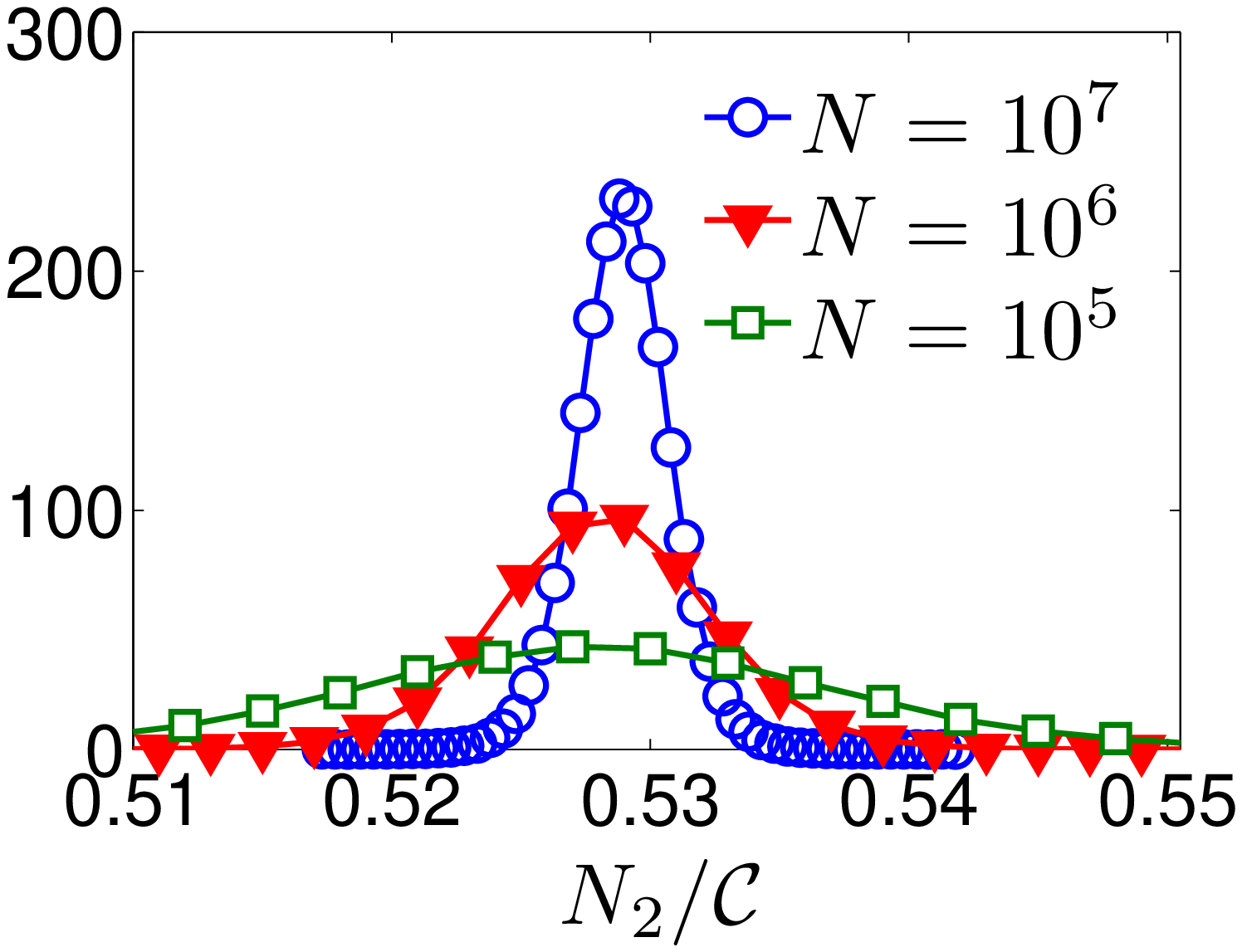}}
\caption{Probability densities in enhanced redirection for (a)
  $\mathcal{C}/N^{\nu\!-\!1}$, $N_2/N^{\nu\!-\!1}$, and $N_3/N^{\nu\!-\!1}$
  for $N=10^6$ (open symbols) and $N=10^7$ nodes (closed symbols) and (b)
  $N_2/\mathcal{C}$.  Data are based on $10^5$ realizations with
  $\lambda=\frac{3}{4}$ and $\nu=1.73$.}
  \label{NkDist}
\end{center}
\end{figure}

In contrast, for enhanced redirection networks, essentially all geometrical
features are non self-averaging, as illustrated by the distributions of
$\mathcal{C}/N^{\nu-1}$, $N_2/N^{\nu-1}$, $N_3/N^{\nu-1}$, etc., which do not
sharpen as $N$ increases (Fig.~\ref{NkDist}).  Since the number of core nodes
$\mathcal{C}$ and the number of nodes of fixed degree $N_k$ for $k\geq 2$ all
scale as $N^{\nu-1}$ (Eq.~\eqref{scaling}), scaled distributions of
$N_k/N^{\nu-1}$ and $\mathcal{C}/N^{\nu-1}$ would progressively sharpen as
$N$ increases if self-averaging holds.  The lack of self-averaging implies a
sensitive dependence on initial conditions where events early in the
evolution have lasting effects on the network structure.  Surprisingly, the
ratios $N_k/\mathcal{C}$ \emph{are} self-averaging for $k\ge 2$, as the
distributions $N_k/\mathcal{C}$ do sharpen as $N$ increases
(Fig.~\ref{NkDist}).  The self-averaging of these ratios suggests that the
degree distributions \emph{given} a value of $\mathcal{C}$ are statistically
similar, even though the overall number of core nodes $\mathcal{C}$ varies
widely between realizations.

\section{Singular Structures}
\label{SS}

Because of the tendency to connect to high-degree nodes, enhanced redirection
networks tend to be dominated by one or a few high-degree nodes.  In this
section, we explore some of the consequences of this attraction to
high-degree nodes.

\subsection{Macrohubs}
\label{macrohubs}

Macrohubs always arise when $\lambda>0$, but they are easily detectable only
when the exponent $\nu$ is notably smaller than 2.  Figure \ref{nu} shows
that this happens when $\lambda \agt 0.4$, and in this range macrohubs are
clearly observed in all network realizations.  When the redirection parameter
is small, $0<\lambda<0.4$, it may be necessary to grow the network to an
astronomically large value of $N$ to detect macrohubs with certainty.

\begin{figure}[ht]
\begin{center}
\subfigure[]{\includegraphics[width=0.45\textwidth]{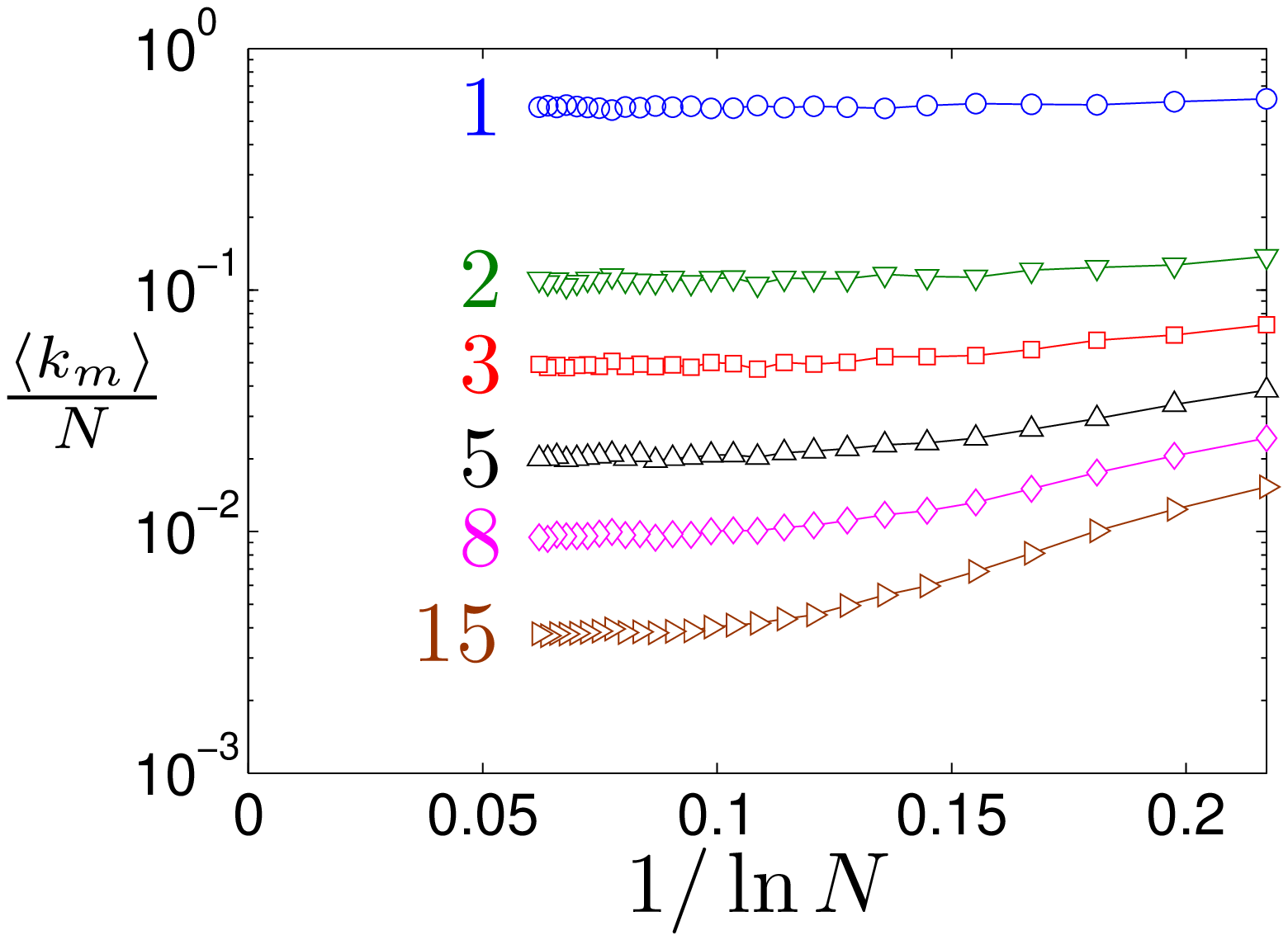}}\qquad
\subfigure[]{\raisebox{0.15 em}{\includegraphics[width=0.45\textwidth]{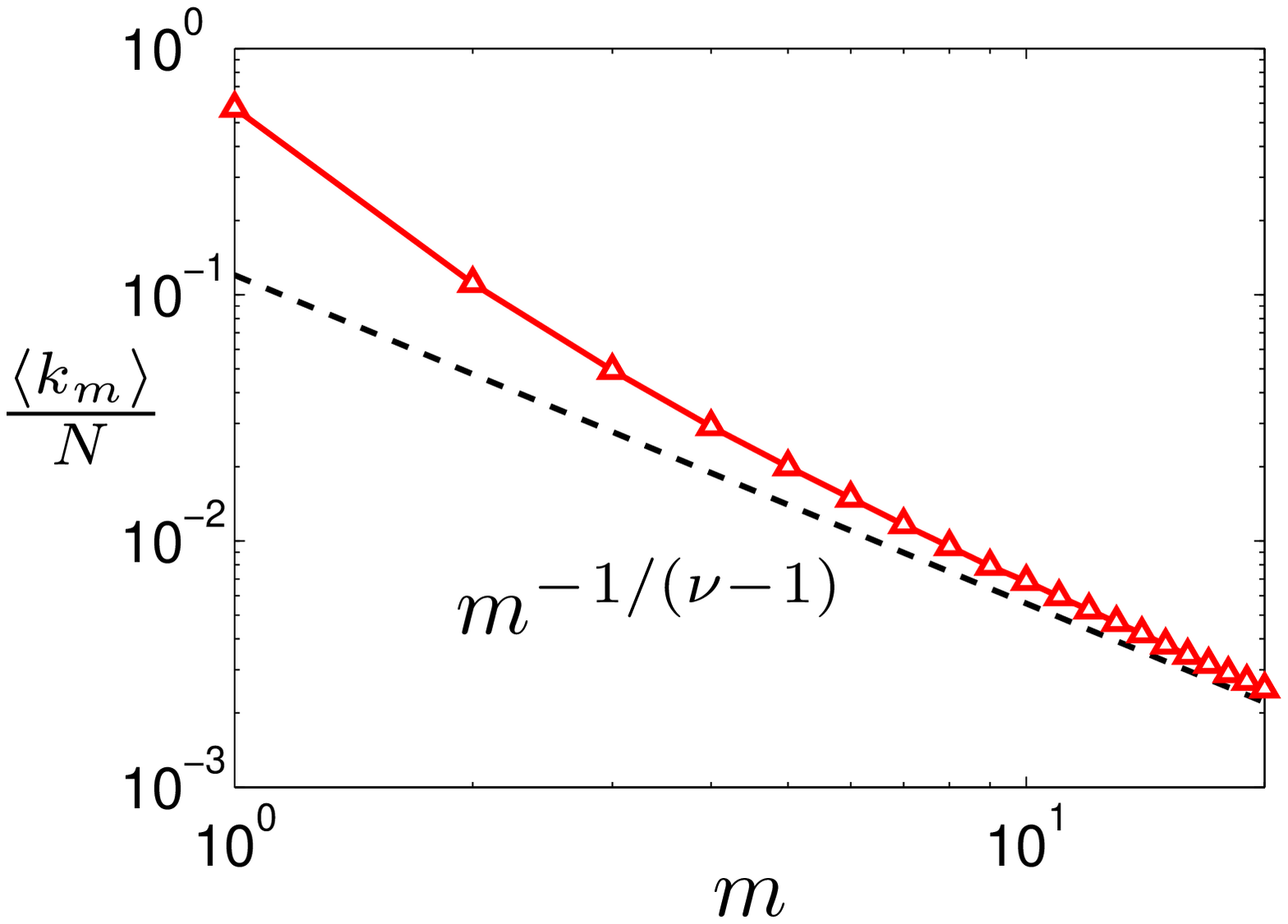}}}
\caption{(a) The averages $\langle k_m\rangle/N$ versus $1/\ln N$ for
  selected $m$, illustrating that these values approach a finite fraction as
  $N\rightarrow\infty$.  Each data point corresponds to $10^3$ realizations.
  (b) $\langle k_m\rangle/N$ versus $m$ for $N=10^6$.  }
  \label{manykm}
\end{center}
\end{figure}

There are usually many macrohubs, whose degree is proportional to $N$, as
shown in Fig.~\ref{manykm}(a).  To estimate $k_m$, the degree of the
$m^\text{th}$ largest macrohub, we use the extremal criterion~\cite{G58},
\begin{equation}
\sum_{k\geq k_m} N_k \sim m\,.
\end{equation}
This equation merely states that there should be of the order of $m$ nodes of
degree $k_m$ or larger.  Thus $k_m$ indeed gives an estimate for the value of
the $m^{\rm th}$-largest degree.  Combining this criterion with the
asymptotic $N_k\sim N^{\nu-1}/k^{\nu}$ from Eq.~\eqref{NkN} gives
\begin{equation}
\label{km}
k_m\sim N/m^{1/(\nu-1)}.
\end{equation}
The basic feature is that the degrees of macrohubs scale \emph{linearly}
with $N$.  In contrast, for networks with an extensive degree distribution of
the form $N_k\sim N/k^{\nu}$ and $\nu>2$, the above extremal criterion gives
the sub-linear growth: $k_{m}\sim (N/m)^{1/(\nu-1)}$.

The degrees of the macrohubs substantially depends on the early stages of
network growth, but once a set of macrohubs emerges (with degrees $k_1$,
$k_2$, $k_3$, $\dots$), the probability of attaching to a macrohub of degree
$k_m$ asymptotically approaches to $k_m/N$.  This preferential attachment to
macrohubs is similar to a P\'olya urn process for filling an urn with balls
of several colors~\cite{EP23,urn}.  In the urn process, a ball is drawn at
random from the urn and then replaced along with an additional matching-color
ball.  The different colors in the urn process correspond to different
macrohubs in enhanced redirection.  If there are $k_m$ balls of the $m^{th}$
color in an urn of $N$ total balls, then the probability of choosing color
$m$, and thus increasing $k_m$, is given by $k_m/N$.  For the P\'olya urn
process, the ultimate fractions of balls of different colors do not
self-average; the same is expected for the scaled degrees of macrohubs in
enhanced redirection.

\subsection{Star Graphs}

Because of the tendency to link to high-degree nodes, it is possible that a
star graph arises in which single node is connected to every other node of
the network.  As we now show, the probability for such a star to occur is
non-zero for $\lambda>1$.  For the initial condition of a single node with a
self loop, the star contains $N-1$ leaves, while the root node has degree
$N+1$.  The probability $S_{N}$ to build such a star graph is
\begin{align}
\label{S_max}
S_{N}(\lambda) = \prod_{n=1}^{N-1}\left\{\frac{1}{n}+\frac{n-1}{n}\left[1 -
    (n+1)^{-\lambda} \right]\right\}.
\end{align}
The factor $\frac{1}{n}$ accounts for the new node attaching to the root in a
network of $n$ nodes, while the second term accounts for first choosing a
leaf and then redirecting to the root.  As shown in~\ref{app:star},
the asymptotic behavior of \eqref{S_max} is
\begin{equation}
\label{S_max_asymp}
S_{N}(\lambda) \to 
\begin{cases}
\mathcal{S}(\lambda)     & \qquad \lambda>1\\
A/N                                 & \qquad \lambda=1\\
\exp\!\left(\!-\frac{N^{1-\lambda}}{1-\lambda}\right)    &\qquad 0<\lambda<1\\
\frac{1}{(N-1)!}    &\qquad  \lambda=0
\end{cases}
\end{equation}
where $\mathcal{S}(\lambda)=\lim_{N\to\infty}S_{N}(\lambda)$ is a
monotonically increasing function of $\lambda$ when $\lambda>1$ and
$A=\pi^{-1}\sinh\pi\approx 3.676$.

\begin{figure}[ht]
\begin{center}
\includegraphics[width=0.45\textwidth]{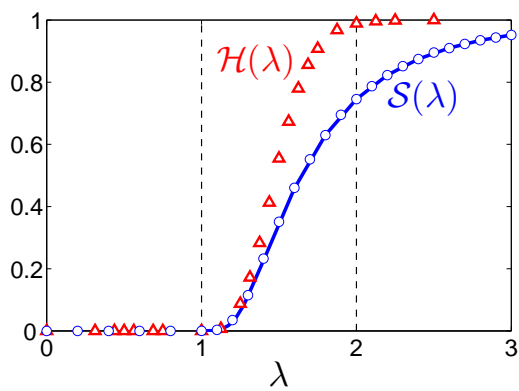}
\caption{Probabilities for a star (\textcolor{blue}{$\circ$}) and a hairball
  graph (\textcolor{red}{$\Delta$}), $\mathcal{S}$ and $\mathcal{H}$,
  respectively.  For each $\lambda$, the data are based on $10^5$
  realizations.  The solid curve represents the numerical evaluation of the
  product in \eqref{S_max}. }
  \label{star}
\end{center}
\end{figure}

For $0<\lambda\leq 1$, the probability of a star graph asymptotically
approaches zero as $N\to\infty$.  In this range, the network typically has
many macrohubs with average sizes distributed according to Eq.~\eqref{km}.
For $\lambda>1$, a star graph occurs with a positive probability
(Fig.~\ref{star}), with a continuous phase transition at $\lambda=1$.  As
shown in \ref{app:star}, this phase transition has an infinite order because all
derivatives of $\mathcal{S}(\lambda)$ vanish at $\lambda=1$.


\subsection{Size Distributions of Macrohubs}

We now study the size distribution of the largest macrohub, the
$2{^\text{nd}}$ largest macrohub, etc.  By `size' we mean the degree of a
macrohub, so that there is no confusion between the size of the largest
macrohub (a quantity characterizing one node) and the degree distribution
which specifies the number of nodes of a fixed degree.

The degree $k_m$ of the $m^\text{th}$ largest macrohub scales linearly with
the total number of nodes $N$, and therefore the corresponding size
distribution $M_m(k_m, N)$ approaches the scaling form
\begin{equation}
\label{hub_scaling}
M_m(k_m, N) \to \frac{1}{N}\,\mathcal{M}_m(x), \quad x = \frac{k_m}{N}
\end{equation}
in the $N\to\infty$ limit.  We do not know how to compute the scaling
functions $\mathcal{M}_m(x)$, but some generic properties of these functions
can be established without calculations.  For instance, the scaling function
$\mathcal{M}_m(x)$ vanishes when $x>1/m$.  Indeed, the $m^\text{th}$-largest
macrohub has maximal degree $k_m=N/m$, which corresponds to the situation
when the first $m$ largest macrohubs all have equal maximally possible size
$N/m$.  Thus $\mathcal{M}_m(x)$ is singular at $x=1/m$:
$\mathcal{M}_m(x)\equiv 0$ when $x>1/m$, and $\mathcal{M}_m(x)>0$ when
$x<1/m$.  Consider now the most interesting function $\mathcal{M}_1(x)$,
which describes the scaled degree of the largest macrohub.  It has a
singularity at $x=1$, and also a singularity at $x=1/2$, as at this point the
second-largest macrohub can emerge.  Continuing this line of reasoning, we
conclude that $\mathcal{M}_1(x)$ has infinitely many singularities that are
located at $x=1,1/2, 1/3, 1/4,\ldots$.  The emergence of these progressively
weaker singularities is a generic feature and they arise in numerous examples
including random walks, random maps, spin glasses, fragmentation, etc., which
are characterized by the lack of self-averaging, see, e.g.,
\cite{DF,Higgs,FIK,DJM,KGB}.  Similarly, the scaling function
$\mathcal{M}_m(x)$ has singularities at $x=1/m, 1/(m+1), 1/(m+2),\ldots$.

The presence of infinitely many singularities (partly) explains why it is
difficult to compute the scaling functions $\mathcal{M}_m(x)$.  Fortunately,
it is possible to probe the asymptotic behavior of $\mathcal{M}_m(x)$ near
the maximal possible size $x=1/m$.  Consider the most interesting case of the
largest macrohub.  To determine the asymptotic behavior of $\mathcal{M}_1(x)$
in the $x\to 1$ limit, we notice that it can be extracted from the
probability to build a star graph.  Comparing \eqref{S_max_asymp} with
\eqref{hub_scaling} we find that in the marginal case of $\lambda=1$
\begin{equation}
\label{M_11}
\mathcal{M}_1(1) = A = \frac{\sinh(\pi)}{\pi}\,, 
\end{equation}
while in the $0<\lambda<1$ range, the scaling function $\mathcal{M}_1(x)$
very rapidly vanishes near the upper limit:
\begin{equation}
\label{M_1_less}
\ln \mathcal{M}_1(x)\sim - (1-x)^{-(1-\lambda)}\qquad\text{when}\quad x\to 1.
\end{equation}
The asymptotic behavior of $\mathcal{M}_2(x)$ in the $x\to \frac{1}{2}$ limit
can be similarly extracted from the probability to build a $2-$star graph. We
outline some of these calculations in the following subsection.

\subsection{Hairballs}

\begin{figure}[ht]
\begin{center}
\subfigure[]{\includegraphics[width=0.4\textwidth]{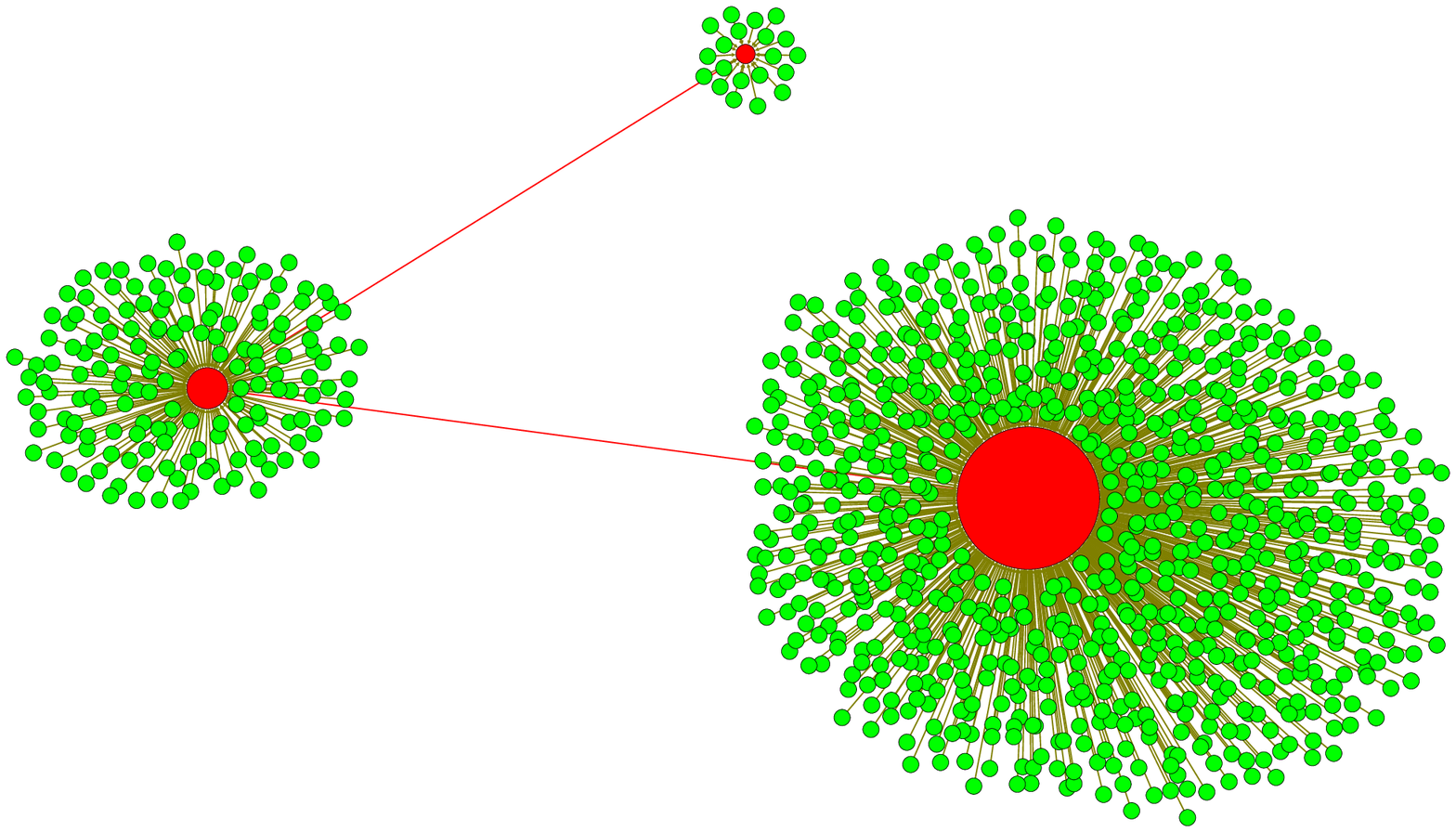}} \qquad
\subfigure[]{\includegraphics[width=0.35\textwidth]{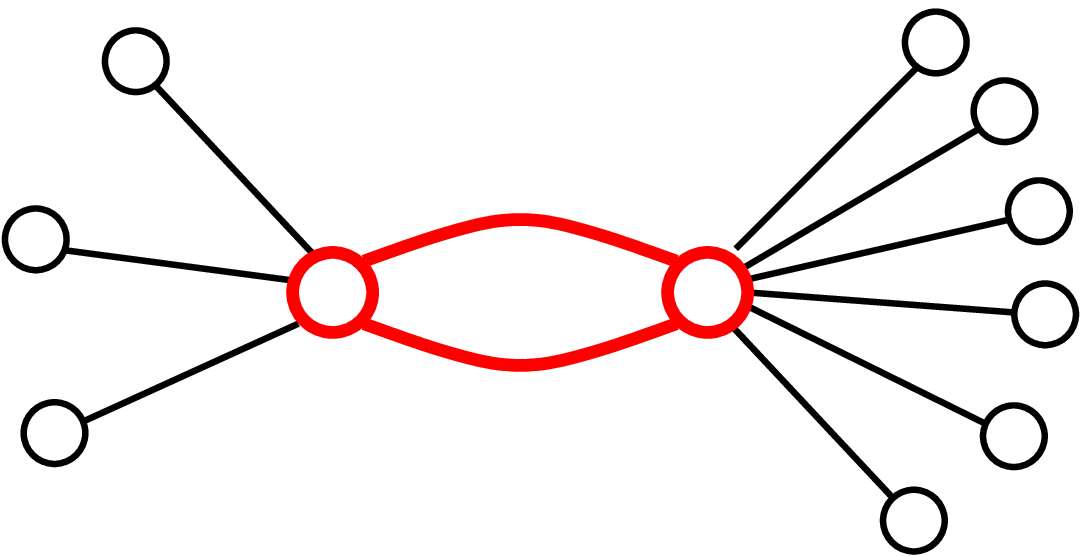}} 
\caption{(a) Example hairball network of $N=10^3$ nodes generated by enhanced
  redirection with $\lambda=2.5$.  There are 3 core nodes (degree $\ge 2$)
  with degrees $k=819$, $163$, and $21$. (b) Idealized hairball graph
  $H_{m,n}$, with $(m,n)=(3,6)$.  The initial nodes and links are
  highlighted. }
  \label{hairball}
\end{center}
\end{figure}

A slightly less singular variant of the star graph is what we term the
``hairball'' graph (Fig.~\ref{hairball}).  A hairball consists of multiple
linked stars in which there are no nodes of degree 2.  A star is thus a
special case of hairball that consists of a single ball. The star probability
is therefore always less than the hairball probability $\mathcal{H}(\lambda)$
(see also Fig.~\ref{star}), and the latter appears to reach 1 for $\lambda> 2$.  Thus
enhanced redirection networks undergo two distinct phase transitions: (i)
emergence of star graph when $\lambda>1$ and (ii) the vanishing of nodes of
degree 2 when $\lambda>2$.

It is difficult to determine the hairball probability $\mathcal{H}(\lambda)$
analytically, as the number of high degree nodes and their degrees are
unspecified.  To gain insight, we consider the more tractable probabilities
for concrete types of hairballs.  Let us start with the simplest hairball
that contains two macrohubs.  We also modify the initial condition to make
the calculations more clean; namely, two nodes in a cycle of size 2
(Fig.~\ref{hairball}(b)).  A hairball with two macrohubs is thus a network
where all nodes, apart from the two initial nodes, are leaves.  Suppose that
the network has reached the stage when one initial node is connected to $m$
leaves and the other initial node is connected to $n$ leaves. Let $H_{m,n}$
be the probability to reach such an $(m,n)$ hairball. This hairball can arise
from an $(m-1,n)$ or an $(m,n-1)$ hairball, which (by definition) occur with
probabilities $H_{m-1,n}$ and $H_{m,n-1}$.  Generally
\begin{equation}
\label{GH}
H_{m,n} = g_{m-1,n}\, H_{m-1,n} + g_{n-1,m}\,H_{m,n-1}\,,
\end{equation}
with coefficients $g$ that depend on the redirection rule.  For redirection
probability $r(a,b)=1-b^{-\lambda}$
\begin{equation}
\label{Gmn_gen}
g_{m-1,n-1}=\frac{1}{m+n}\left[m-\frac{m}{(m+1)^\lambda}+\frac{1}{(n+1)^\lambda}\right]~.
\end{equation}
Instead of simulating enhanced redirection networks and looking for
hairballs, we can use the recurrence \eqref{GH} to calculate the exact values
$H_{m,n}$ for any $(m,n)$ starting from the obvious initial condition
$H_{0,0}=1$.

The recurrences \eqref{GH} are readily solvable in the limit when either $m$
or $n$ vanishes.  In this case
\begin{equation}
\label{HNN}
H_{N-2,0}=H_{0,N-2} = \frac{1}{2}\prod_{n=1}^{N-2}\left[1-\frac{n-1}{n^{\lambda+1}}\right]
\end{equation}
for $N\geq 3$. Hence the probability to generate the star graph, $S_N=H_{N-2,0}+H_{0,N-2}$, is 
\begin{equation*}
S_N(\lambda)=\prod_{n=1}^{N-2}\left[1-\frac{n-1}{n^{\lambda+1}}\right]~.
\end{equation*}
Using the same analysis as that used to derive Eq.~\eqref{S_max_asymp}, we
find the asymptotic behaviors
\begin{equation}
\label{S_max_2}
S_{N}(\lambda) \to 
\begin{cases}
\mathcal{S}(\lambda)     & \qquad \lambda>1\,,\\
B/N                                 & \qquad \lambda=1\,,\\
\exp\!\left(-\frac{N^{1-\lambda}}{1-\lambda}\right)    &\qquad 0<\lambda<1\,,
\end{cases}
\end{equation}
with $\mathcal{S}(\lambda)=\prod_{n\geq 1} \left[1-(n-1)/n^{\lambda+1}\right]$
and
\begin{equation*}
B=\prod_{n=1}^\infty\left[1+\frac{1}{n(n+1)} \right] 
= \frac{1}{\pi}\cosh\!\left(\frac{\pi\sqrt{3}}{2}\right)
= 2.428189792\ldots
\end{equation*}
The differences between \eqref{S_max_asymp} and the above formulae stem from
the different initial conditions.

The most interesting behavior arises when both $m$ and $n$ are large and
comparable: $m\sim n\sim N$.  In this regime we employ a continuum
approach. We treat $m$ and $n$ as continuous variables and expand $H_{m-1,n}$
and $H_{m,n-1}$ in Taylor series to lowest order
\begin{equation}
\label{HHHH}
H_{m-1,n} = H - \frac{\partial H}{\partial m}\,, \qquad
H_{m,n-1} = H - \frac{\partial H}{\partial n}\,,
\end{equation}
where $H\equiv H_{m,n}$.  Substituting the expansions \eqref{HHHH} into
\eqref{GH} and using \eqref{Gmn_gen}, we recast the original recurrence into
a partial differential equation that depends on $\lambda$.  When
$0<\lambda<1$, we obtain
\begin{equation}
\label{H_minus}
m\,\frac{\partial H}{\partial m} + n\,\frac{\partial H}{\partial n}= -\big(m^{1-\lambda}+n^{1-\lambda}\big) H\,.
\end{equation}
The controlling factor of the solution is given by
\begin{equation}
\label{H_minus_sol}
H_{m,n} \sim \exp\!\left[-\frac{m^{1-\lambda}+n^{1-\lambda}}{1-\lambda}\right], \qquad 0<\lambda<1\,.
\end{equation}
To find the sub-leading factors, it would be necessary to refine
\eqref{H_minus} by keeping lower-order terms.

In the marginal case $\lambda=1$, the partial differential equation becomes
\begin{equation}
\label{H_crit}
m\,\frac{\partial H}{\partial m} + n\,\frac{\partial H}{\partial n}= - 3H\,,
\end{equation}
whose solution, which satisfies the necessary symmetry requirement
$H_{m,n}=H_{n,m}$, is
\begin{equation}
\label{H_crit_sol}
H_{m,n} = \frac{C}{(mn)^{3/2}}
\end{equation}
with some amplitude $C$ that cannot be computed in the framework of the
continuum approximation.

Setting $m=n=N/2$ (so that the corresponding total number of nodes in the
network is $N+2$) we see that the probability of such $(N/2,N/2)$ graph
scales as $N^{-3}$.  This is precisely the probability that the
second-largest macrohub has the maximal possible size $N/2$.  The scaling
behavior \eqref{hub_scaling} of the size distribution of the second-largest
hub is compatible with the $N^{-3}$ extremal behavior if
\begin{equation}
\label{M21}
\mathcal{M}_2(x)\sim \left(\tfrac{1}{2}-x\right)^2 \qquad\text{when}\quad  x\to \tfrac{1}{2}\,.
\end{equation}

For $\lambda>1$, the partial differential equation becomes
\begin{equation}
\label{H_plus}
m\,\frac{\partial H}{\partial m} + n\,\frac{\partial H}{\partial n}= - H\,.
\end{equation}
The remarkable feature of this equation is its universality (independence of
$\lambda$) for large $m$ and $n$.  Solving \eqref{H_plus} we get
\begin{equation}
\label{H_plus_sol}
H_{m,n} = \frac{C_2(\lambda)}{\sqrt{m n}}\,, \qquad \lambda>1\,.
\end{equation}
The multiplicative constant factor $C_2(\lambda)$ cannot be determined within
the continuum framework.  Setting again $m=n=N/2$ we find that the second
largest macrohub has the maximal possible size $N/2$ with probability
$2C_2(\lambda)/N$, which in conjunction with the scaling behavior
\eqref{hub_scaling} tells us that
\begin{equation}
\label{M22}
\mathcal{M}_2\big(\tfrac{1}{2}\big)=2C_2(\lambda) \qquad\text{when}\quad \lambda>1\,.
\end{equation}

Let us briefly discuss the general case of a hairball with $p$ macrohubs.  To
simplify the analysis we again modify the initial condition by taking the
initial network to be a cycle of $p$ nodes.  Generalizing the above analysis,
we find that in the marginal case of $\lambda=1$, the governing equation is
\begin{equation}
\label{H_p_crit}
m_1\,\frac{\partial H}{\partial m_1} +\ldots+ m_p\,\frac{\partial H}{\partial m_p}= - (2p-1)H\,,
\end{equation}
whose solution is
\begin{equation}
\label{H_p_sol_crit}
H_{m_1,\ldots,m_p} = C_p\left(\prod_{j=1}^p m_j\right)^{-2+1/p}~.
\end{equation}
When $\lambda>1$, the governing equation for $H$ is
\begin{equation}
\label{H_p}
m_1\,\frac{\partial H}{\partial m_1} +\ldots+ m_p\,\frac{\partial H}{\partial m_p}= - (p-1)H\,,
\end{equation}
whose solution is
\begin{equation}
\label{H_p_sol}
H_{m_1,\ldots,m_p} = C_p(\lambda)\left(\prod_{j=1}^p m_j\right)^{-1+1/p}\,.
\end{equation}
Using these results we extract the asymptotic behavior of the scaled size
distribution $\mathcal{M}_p(x)$ of the $p^\text{th}$ largest macrohub in the
$x\to \tfrac{1}{p}$ limit:
\begin{equation}
\label{Mp}
\mathcal{M}_p(x)\sim 
\begin{cases}
\big(\tfrac{1}{p}-x\big)^{2p-2} & \lambda=1,\\
\big(\tfrac{1}{p}-x\big)^{p-2}  & \lambda>1.
\end{cases}
\end{equation}
Finally when $\lambda<1$, the asymptotic behavior of the scaled size
distribution $\mathcal{M}_p(x)$ is extracted from \eqref{H_minus_sol}, and
its generalization to an arbitrary $p$, to give
\begin{equation}
\ln \mathcal{M}_p(x)\sim - \big(\tfrac{1}{p}-x\big)^{-(1-\lambda)}.
\end{equation}


\subsection{Root Node}

\begin{figure}[ht]
\begin{center}
{\includegraphics[width=0.55\textwidth]{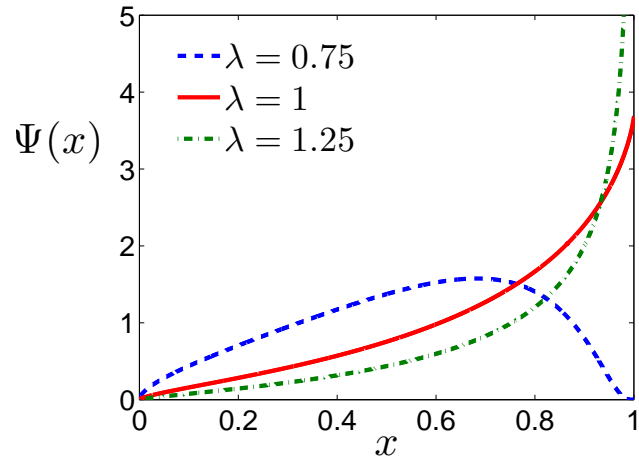}}
\caption{The scaling function from Eq.~\eqref{scaling} for the root degree
  probability distribution. }
  \label{rootNode}
\end{center}
\end{figure}

To further appreciate the role of macrohubs let us now consider the evolution
of the degree of the root node.  We return to our default initial condition
of a single root node that is linked to itself.  Let $R(k,N)$ be the
probability that the root has degree $k$ in a network of $N$ nodes.  This
probability was previously determined analytically for random attachment and
linear preferential attachment networks~\cite{KR02_b}, where it was shown
that the root degree is broadly distributed for preferential attachment.  For
enhanced redirection, this probability obeys the difference equation
\begin{eqnarray}
\label{root}
R(k,N+1) &=& \left\{\!\frac{1}{N}+\frac{k-3}{N}\left[1-\frac{1}{(k-1)^\lambda}\right]\!\right\}R(k-1,N) \nonumber\\
&+& \left\{\!1-\frac{1}{N}-\frac{k-2}{N}\left[1-\frac{1}{k^\lambda}\right]\!\right\}R(k,N)
\end{eqnarray}
with initial condition $R(2,1)=1$.  The first term gives the probability that
the root has degree $k-1$ when the $(N+1)^{\rm th}$ node connects to it,
either directly or by redirection from the $k-3$ children of the root.  The
second term is the probability that the root has degree $k$ and the
$(N+1)^{\rm th}$ node does \emph{not} connect to the root.  Iterating
\eqref{root} numerically we obtain the numerical results for $R(k,N)$ shown
in Fig.~\ref{rootNode}.
 
The behavior of the moments $\langle k^d \rangle_N = \sum_k k^d R(k,N)$ helps
to shed light on the behavior of the root degree distribution $R(k,N)$.  From
Eq.~\eqref{root}, the mean degree evolves according to
\begin{equation}
\label{meanK}
N\langle k \rangle_{N+1}=(N+1)\langle k\rangle_N
-\langle k^{1-\lambda}\rangle_N+2\langle k^{-\lambda}\rangle_N-1 
\end{equation}
While this recurrence is not closed, we can drop the second term on the
right-hand side of \eqref{meanK} as $N\to\infty$ because $\langle
k^{1-\lambda}\rangle_N\ll \langle k\rangle_N$ (since $\lambda>0$).  The
following terms are even smaller.  Hence \eqref{meanK} simplifies to $\langle
k\rangle_{N+1}\simeq \left(1+\tfrac{1}{N}\right)\langle k\rangle_N$, from
which $\langle k\rangle\sim N$.  Similarly, the recurrence for the variance
$\sigma_N=\langle k^2\rangle_N - \langle k\rangle_N^2$ shows that $\langle
\sigma\rangle_{N+1}\simeq \left(1+\tfrac{2}{N}\right)\langle
\sigma\rangle_N$.  This implies that $\langle \sigma\rangle\sim N^2$.  Thus
we conclude that the degree of the root is {\em not} self-averaging.  The
lack of self-averaging arises because the early evolution steps play a huge
role in determining the root degree for large $N$.

Since the root degree is non self-averaging with $\langle k\rangle\sim N$, we
anticipate that when $k,N\to\infty$, the probability distribution $R(k,N)$
admits the scaling form (Fig.~\ref{rootNode})
\begin{equation}
\label{scaling}
R(k,N) = N^{-1}\Psi(x), \qquad x=\frac{k}{N}
\end{equation}
When $\lambda\leq 1$, the scaled distribution $\Psi(x)$ is a smooth function
on $[0,1]$, but when $\lambda>1$, $\Psi(x)$ additionally contains a singular
component $\mathcal{R}(\lambda)\delta(x-1)$ that accounts for the probability to
create a star or a hairball about the root node.  Therefore
\begin{equation}
\label{RS}
\mathcal{R}(\lambda) \geq \mathcal{S}(\lambda) = \prod_{n=1}^\infty \left[1 - \frac{n-1}{n(n+1)^\lambda} \right].
\end{equation}

Consider the extreme behaviors at $x\to 1$. This limit essentially
corresponds to the probability of forming the star. Thus for $\lambda=1$ we
expect $\Psi(1)=\pi^{-1}\sinh\pi\approx 3.676$. This agrees with simulation
results. Further, $\Psi(1)=0$ for $\lambda<1$; more precisely, the scaling
function $\Psi(x)$ near $x=1$ is essentially the same as the scaling function
$\mathcal{M}_1(x)$ describing the largest macrohub,
Eq.~\eqref{M_1_less}. Hence $\ln\Psi(x) \sim -(1-x)^{-(1-\lambda)}$ as $x\to
1$. This is also compatible with our numerical results.

To estimate the asymptotic behavior of $\Psi(x)$ for $x\to 0$, we have
computed the probabilities that the root has the smallest possible degrees
$k=3$ and $k=4$ (\ref{app:small}).  From these results we infer the
asymptotic behavior
\begin{equation}
\label{Psi_small}
\Psi(x)\sim x^{1-3^{-\lambda}}\qquad\text{as}\quad x\to 0\,,
\end{equation}
which is compatible with our numerical results.

\section{Enhanced Redirection with Multiple Attachments}
\label{DA}

In our discussion thus far, each new node added to the network has a single
outgoing link. The resulting network is therefore a tree, except for closed
loops that were part of the initial condition.  It is therefore worthwhile to
check whether the many anomalous features of enhanced redirection still exist
if we allow the out-degree of each node to be larger than 1 so that closed
loops can be created.

\begin{figure}[ht]
\begin{center}
{\includegraphics[width=0.4\textwidth]{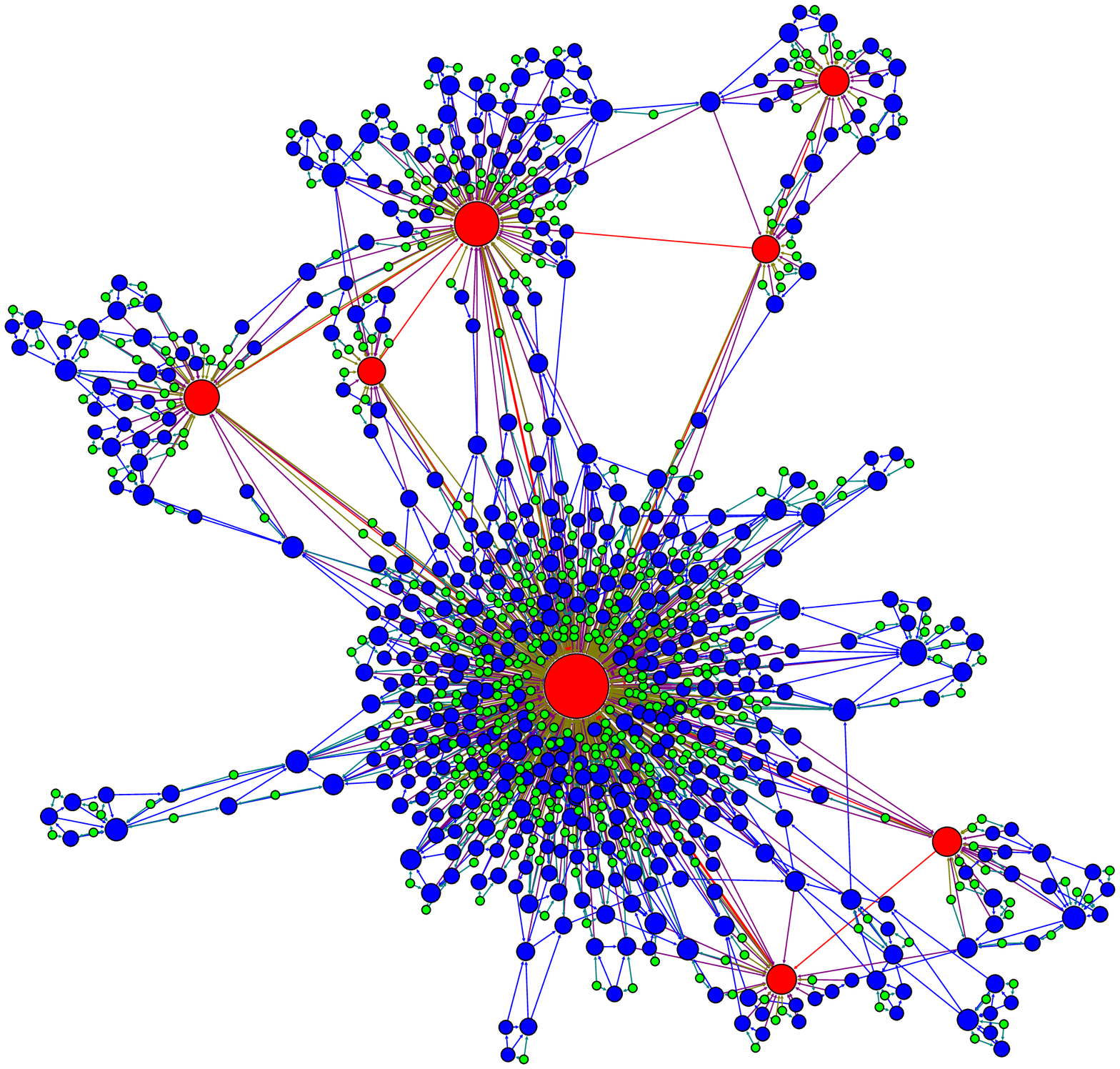}}
\caption{A network of $N=10^3$ nodes which has been built by enhanced
  redirection with double attachment, Eq.~\eqref{r12} with $\lambda=0.75$.
  The maximum degree in this example is $k_{\rm max}=623$.}
  \label{double}
\end{center}
\end{figure}

For simplicity, we consider the attachment rule in which a new node makes
exactly two connections to existing nodes of the network---double attachment.
We choose the initial condition of a single node with two self-loops, so that
the root node is its own parents.  Nodes are added sequentially according to
the following rules:
\begin{enumerate}
\item The new node links to a randomly selected target node.
\item The new node links to one of the two parents of the target with
  probabilities
\begin{equation}
\label{r12}
r_1(a,b)=\frac{a^\lambda}{a^\lambda+b^\lambda}, \qquad 
r_2(a,b)=\frac{b^\lambda}{a^\lambda+b^\lambda},
\end{equation}
where where $a$ and $b$ are the degrees of parent 1 and parent 2.
\end{enumerate}
We choose these redirection probabilities so that attachment to a given
parent becomes increasingly likely as its degree increases:
$r_1(a,b)\rightarrow 1$ as $a\rightarrow\infty$ and $r_2(a,b)\rightarrow1$ as
$b\rightarrow\infty$.  Figure~\ref{double} shows a typical network that has
been generated by rules (i) and (ii) with $\lambda=0.75$.


\subsection{Degree Distribution}

\begin{figure}[ht]
\begin{center}
\includegraphics[width=0.5\textwidth]{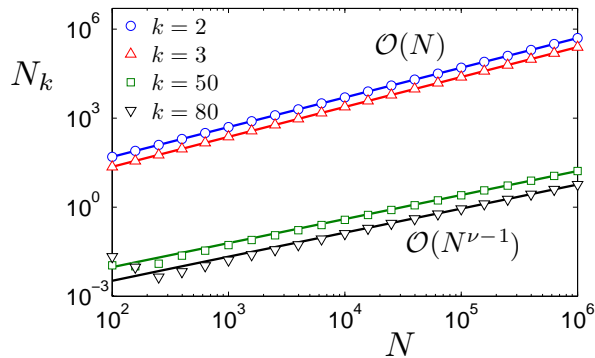}
\caption{$N_k$ versus $N$ for degrees $k=2$, $3$, and degrees $k=50$, $80$
  for $\lambda=0.75$.  The lines indicate the scaling $N_k\sim N$ and
  $N_k\sim N^{\nu-1}$ with $\nu=1.8$. }
  \label{NkDouble}
\end{center}
\end{figure}

The two links created by each new node arise from two qualitatively different
mechanisms.  One link arises by random attachment, a mechanism that leads to
the random recursive tree.  The second link is to one of the two parents, and
it is selected via enhanced redirection which, as we have shown earlier,
leads to a broad and non-extensive degree distribution, $N_k\sim
N^{\nu-1}/k^{\nu}$ with $\nu<2$.  Because of the competition between these
two mechanisms, one may anticipate that $N_k$ scales linearly with $N$ for
sufficiently small $k$, while for large $k$ the degree distribution $N_k$
scales as $N^{\nu-1}$ with $\nu<2$. This is indeed the case, as illustrated
in Fig.~\ref{NkDouble}.

To understand how two different scaling regimes emerge, we study the master
equation that governs $N_k$ (compare with Eq.~\eqref{master1}):
\begin{equation}
\label{masterDouble}
\frac{dN_k}{dN}=\left[\frac{N_{k-1}}{N}-\frac{N_k}{N}\right]
+\left[\frac{\tau_{k-1}(k-3)N_{k-1}}{N}-\frac{\tau_{k}(k-2)N_{k}}{N}\right]
+\delta_{k,2}\,.
\end{equation}
The terms in the first set of brackets account for attachment to a
randomly-selected target.  Similarly, the terms in the second set of brackets
account for redirection.  Here
\begin{equation}
\label{tau}
\tau_k=\sum_b \frac{r_1(k,b){N}(k,b)}{(k-2)N_k},
\end{equation}
is the probability that an incoming node attaches to the degree-$k$ parent of
a random target, where ${N}(a,b)$ is the number of nodes with parents of
degree $a$ and $b$.  Thus $\tau_k$ is the probability of redirection to a
degree-$k$ parent averaged over all $(k-2)N_k$ children of this parent node.
In \eqref{masterDouble}, the expression $\tau_{k}(k-2)N_{k}/N$ gives the
probability that the incoming node initially targets one of the $(k-2)N_k$
children of a node of degree $k$ and then redirects to this parent.  The term
$\delta_{k,2}$ accounts for each newly-created node has degree $2$.

To determine $N_k$, we separately analyze Eq.~\eqref{masterDouble} for small
and for large $k$.  When $k$ is small, we make the ansatz $N_k=c_kN$ and
substitute into Eq.~\eqref{masterDouble}.  Rearranging, we find the recursion
relation for $c_k$ for $k>2$:
\begin{equation}
c_k=\frac{1+(k-3)\tau_{k-1}}{2+(k-2)\tau_k}c_{k-1}\,,
\end{equation} 
while $c_2=\frac{1}{2}$.  This recursion has the product solution
\begin{equation}
\label{prodSmallK}
c_k=\frac{1}{2^{k-1}}\prod_{j=3}^k\frac{1+(j-3)\tau_{j-1}}{1+(j-2)\tau_j/2}~.
\end{equation}
Since redirection to a low degree parent is unlikely, we approximate the
redirection probability as $\tau_k=0$ for small $k$.  With this
approximation, the product in Eq.~\eqref{prodSmallK} equals to $1$, and the
degree distribution reduces to $N_k\approx N/2^{(k-1)}$.

For large $k$, we substitute the non-extensive scaling ansatz
$N_k=c_kN^{\nu-1}$ into Eq.~\eqref{masterDouble}, and rearrange to obtain the
recursion
\begin{equation}
c_k=\frac{1+(k-3)\tau_{k-1}}{\nu+(k-2)\tau_k}c_{k-1}\,.
\end{equation}
This gives the product solution
\begin{equation}
\label{prodLarge}
c_k=c_{\ell}\prod_{j=\ell+1}^k\frac{1+(j-3)\tau_{j-1}}{\nu+(j-2)\tau_j},
\end{equation}
where $\ell$ is the degree above which the non-extensive scaling ansatz is
valid.  In the limit of large $k$, we approximate $\tau_k=1$ in
Eq.~\eqref{prodLarge}because the probability of redirection to a high-degree
parent node approaches 1.  This gives the asymptotic behavior
\begin{equation}
c_k=c_\ell\prod_{j=\ell+1}^k\frac{j-2}{j-2+\nu}\sim k^{-\nu}.
\end{equation}
Combined with the non-extensive ansatz, the degree distribution for large $k$
is $N_k\sim N^{\nu-1}/k^{\nu}$.  In the above derivation, the precise value
of $\ell$ only affects $c_k$ up to a multiplicative factor but not the
scaling behavior at large $k$.

\begin{figure}[ht]
\begin{center}
\subfigure[]{\includegraphics[width=0.45\textwidth]{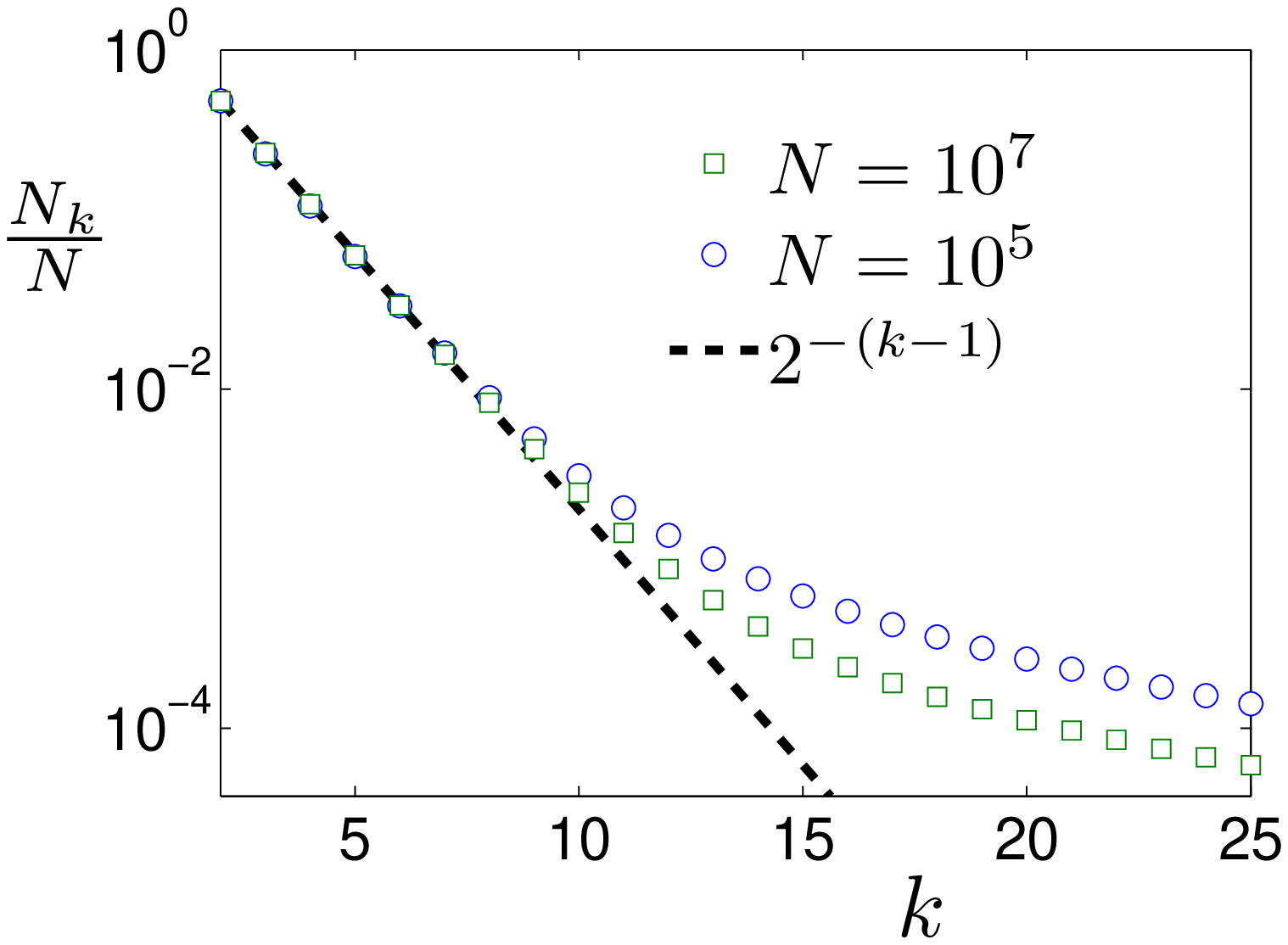}}\quad
\subfigure[]{\includegraphics[width=0.45\textwidth]{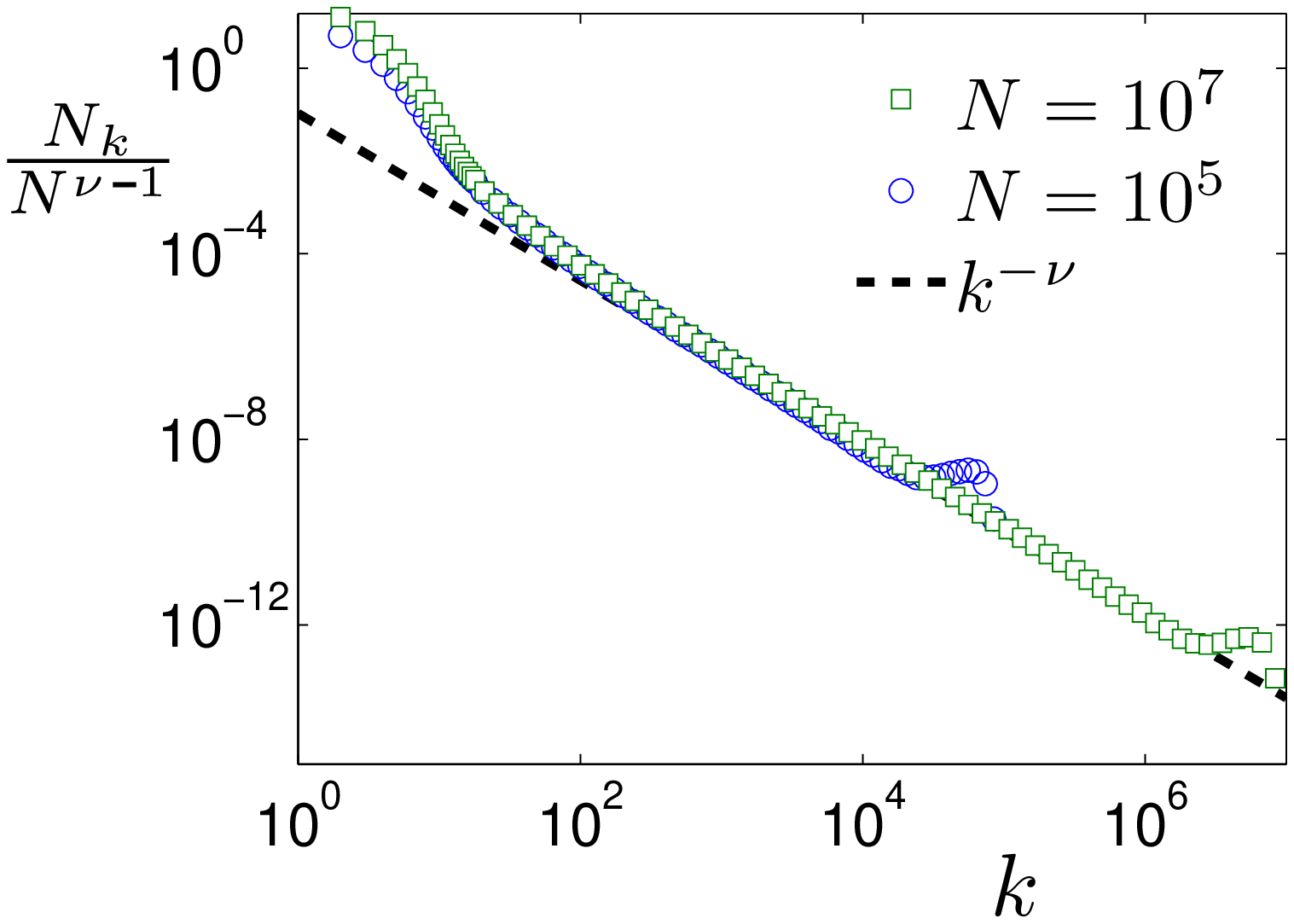}}
\caption{The degree distribution for double attachment with $\lambda=0.75$
  and $\nu=1.8$ for $N=10^5$, $10^6$, and $10^7$.  In (a), the scaled degree
  distribution $N_k/N$ is plotted to show the data collapse for small $k$.
  Similarly, (b) plots $N_k/N^{\nu-1}$ show the data collapse at large $k$.}
  \label{NkDistDouble}
\end{center}
\end{figure}
We now define the crossover degree $k^*$ as the value that separates the
small-$k$ extensive scaling regime from the large-$k$ non-extensive regime.
To estimate $k^*$, we find the value at which $N_k$ in the small- and
large-$k$ approximations coincide.  This leads to the transcendental equation
$N/2^{(k^*-1)}= N^{\nu-1}/(k^*)^{\nu}$, whose solution gives $k^*\sim \ln N$
to lowest order in $N$.

To summarize, the limiting degree distributions are
\begin{equation}
\label{NkSolveD}
N_k=\begin{cases}
N/2^{(k-1)} &\qquad  k\ll k^*\,, \\
cN^{\nu-1}/k^{\nu} &\qquad k\gg k^*\,,
\end{cases}
\end{equation}
with $c$ a constant.  As shown in Fig.~\ref{NkDistDouble}, the agreement
between this prediction and the degree distribution from Monte Carlo
simulations is excellent.  The double attachment rule also produces many
macrohubs whose degrees grow linearly with $N$ and also obey the same scaling
behavior \eqref{km} as in single-attachment enhanced redirection.


\subsection{Clustering Coefficient}

The new feature of double attachment is that the resulting network contains
closed loops.  A basic question about this type of network is whether it is
homogeneous or highly clustered.  We measure the level of clustering by the
\emph{local clustering coefficient} $C_i(k)$~\cite{N10} for a given node $i$
of degree $k$.  This quantity is defined as the ratio of the actual number of
links between the neighbors of node $i$ to the $k(k-1)/2$ possible links
between these neighbors if they were all connected.  For the complete graph,
the local clustering coefficient equals 1 for every node, while in a tree
network the local clustering coefficient is everywhere zero.

To compute the local clustering coefficient for double attachment networks,
consider an arbitrary node $i$ with degree $k$.  There are two ways that
attachment can occur to this node: (i) a new node can attach directly to node
$i$ and to one of its parents, or (ii) a new node can attach to one of the
$k-2$ children of node $i$ and to node $i$ itself.  In either case, the
degree of node $i$ increases by one and the number of links between the
neighbors of node $i$ also increase by one.  When node $i$ is first created
it necessarily has degree 2 and a single link between its neighbors.  Thus
when node $i$ reaches degree $k$, it will have $k-1$ links between its
neighbors.  Therefore its clustering coefficient is
$C_i(k)=(k-1)/[k(k-1)/2]=2/k$.  This $1/k$ scaling of the local clustering
coefficient is seen in many real-world networks \cite{WS98,ABCG06,FLW08,RB03}.

The average local clustering coefficient is therefore
\begin{subequations}
\begin{equation}
\langle C_i \rangle = \frac{1}{N}\sum_{k=2}^{2N}C(k)N_k=\frac{2}{N}\sum_{k=2}^{2N}\frac{N_k}{k}~.
\end{equation}
We partition the sum according to whether $k$ is smaller or greater than
$k^*$ using Eq.~\eqref{NkSolveD}.  This gives
\begin{equation}
\langle C_i \rangle = \sum_{k=2}^{k^*}\frac{1}{2^{(k-2)}\,k}\, +\, N^{\nu-2}\!\sum_{k=k^*+1}^{2N}\frac{2}{k^{\nu+1}}\,,
\end{equation}
with $k^*\sim \ln N$ the cutoff between the two scaling regimes.  As
$N\rightarrow\infty$, the second term vanishes, because $\nu<2$, and the
first term asymptotically approaches
\begin{equation}
\label{cluster}
\langle C_i \rangle \rightarrow \sum_{k=2}^\infty 2^{-(k-2)}k^{-1}=4\ln2-2=0.77258\dots
\end{equation}
\end{subequations}
This large value for this coefficient value indicates a highly clustered
network.  For comparison, empirical studies found $\langle C_i \rangle =
0.79$ for actor collaboration networks~\cite{WS98}, $0.68$ for co-authorship
networks~\cite{ABCG06}, and $0.14$ for blogging networks~\cite{FLW08}.  By
contrast, the Erd\H os-R\'enyi random graphs have a vanishing mean local
clustering coefficient; more precisely it decreases with $N$ according to
$\langle C_i\rangle\sim N^{-1}$ \cite{RB03}.

\section{Conclusion}

Enhanced redirection is an appealing mechanism that produces
networks with a variety of unusual features, including the existence of
multiple macroscopic hubs, anomalous scaling of the degree distribution, and
lack of self-averaging.  Unlike other models that produce macrohubs,
enhanced redirection is based solely on local growth rules and does not
assume intrinsic differences between nodes.  Networks grown by enhanced
redirection are highly disperse and typically consist of a set of
loosely-connected macrohubs that is reminiscent of airline
route networks~\cite{N10,CS03,BO99,H04,GMTA05}.

Intriguingly, the degree distribution decays more slowly than $k^{-2}$.  Such
an anomalously slow decay is mathematically consistent with a finite average
degree only if the number of nodes of fixed degrees scales sub-linearly with
the number of nodes $N$.  Enhanced redirection may thus provide the mechanism
that underlies the wide range of networks~\cite{KBM13} whose degree
distributions apparently decay more slowly than $k^{-2}$.


We also combined the enhanced redirection mechanism with the simplest random
attachment to produce networks that contain closed loops.  The resulting
degree distribution exhibits an unusual combination of extensive and
non-extensive scaling.  The clustering coefficient in these networks is
large, as is observed in many real networks, and thus this rule produces
highly-clustered network with numerous macrohubs.

\bigskip\noindent This research was partially supported by the AFOSR and
DARPA under grant \#FA9550-12-1-0391 and by NSF grant No.\ DMR-1205797.

\appendix
\section{Star Probability}
\label{app:star}

To derive the asymptotic behaviors given in \eqref{S_max_asymp} we first
re-write \eqref{S_max} as
\begin{equation}
\label{S_maximum}
S_{N}(\lambda) = \prod_{n=1}^{N-1}\left[1 - \frac{n-1}{n(n+1)^\lambda} \right].
\end{equation}
When $\lambda>1$, the product on the right-hand side of \eqref{S_maximum} converges to 
\begin{equation}
\label{S_inf}
\mathcal{S}(\lambda) = \prod_{n=1}^\infty \left[1 - \frac{n-1}{n(n+1)^\lambda} \right].
\end{equation}
The probability $\mathcal{S}(\lambda)$ is clearly positive and an increasing
function of $\lambda>1$ (see also Fig.~\ref{star}).  Numerical evaluation of
the product gives, for example, $\mathcal{S}(2) \approx 0.74562, ~\mathcal{S}(4)
\approx 0.9884, ~\mathcal{S}(6) \approx 0.999$.

When $\lambda\leq 1$, the product on the right-hand side of \eqref{S_maximum}
converges to zero as $N\to\infty$.  Consider first the case where
$0<\lambda<1$.  We take the logarithm of \eqref{S_maximum} and expand the
logarithm.  Since the dominant contribution arises for large $n$, we can
replace the sum by an integral to yield
\begin{eqnarray*}
\ln S_N(\lambda) = \sum_{n=1}^{N-1}\ln\left[1 - \frac{n-1}{n(n+1)^\lambda} \right]
                     \sim -\int^N \frac{dn}{n^\lambda} = - \frac{N^{1-\lambda}}{1-\lambda}~.
\end{eqnarray*}

In the marginal case of $\lambda=1$ we use $\prod_{1\leq n\leq N-1}\left[
  1-(n+1)^{-1}\right] = N^{-1}$ to rewrite Eq.~\eqref{S_maximum} at
$\lambda=1$ as
\begin{equation}
\label{NSN}
NS_N(1)=\prod_{n=1}^{N-1} \frac{ 1-\frac{n-1}{n(n+1)}}{1-\frac{1}{n+1}}=\prod_{n=1}^{N-1}\left[ 1+n^{-2}\right].
\end{equation}
The last product converges to $A=\pi^{-1}\sinh \pi$ as
$N\to\infty$~\cite{AS72}.  In the extreme case of $\lambda=0$, we simplify
Eq.~\eqref{S_max} to give $S_N(0)=1/(N-1)!$.  These give the results
summarized in Eq.~\eqref{S_max_asymp}.  It is worth emphasizing that the
third line in \eqref{S_max_asymp}, $\exp[-N^{1-\lambda}/(1-\lambda)]$
represents only the controlling factor in the asymptotic behavior.
Subdominant, and possibly less singular contributions as $N\to\infty$ have
been neglected.

The precise behavior of $\mathcal{S}(\lambda)$ in the $\lambda-1\to 0^+$
limit can be extracted from \eqref{S_inf}.  Taking the logarithm of the
infinite product for $\mathcal{S}(\lambda)$, expanding the logarithm, and
separating the terms that converge and diverge as $\lambda\downarrow 1$, one
gets
\begin{equation*}
\ln \mathcal{S}(\lambda) = -\zeta(\lambda)+2+
\sum_{n\geq 1}\left(\tfrac{n}{(n+1)(n+2)}+\ln\!\left[1-\tfrac{n}{(n+1)(n+2)}\right]\right)
+\mathcal{O}(\lambda-1)\,.
\end{equation*}
Recalling the asymptotic behavior of the zeta function, $\zeta(\lambda) =
(\lambda-1)^{-1} + \gamma +\mathcal{O}(\lambda-1)$, where $\gamma\approx
0.5772$ is Euler's constant, we conclude that
\begin{equation}
\label{S_inf_asymp}
\mathcal{S}(\lambda) \simeq \exp\!\Big(-\frac{1}{\lambda-1} + C\Big)\,,
\end{equation}
with  
\begin{equation*}
C = 2-\gamma 
+ \sum_{n\geq 1}\left[\tfrac{n}{(n+1)(n+2)}+\ln\!\left(1-\tfrac{n}{(n+1)(n+2)}\right)\right]\approx 1.3018\,.
\end{equation*}

Let us now compute the probability to form the star for the model with
redirection probability $r(a,b)=a^\lambda/(a^\lambda+b^\lambda)$. One gets
\begin{equation*}
S_{N}(\lambda) = \prod_{n=1}^{N-1}\left[1 - \frac{n-1}{n}\,\frac{1}{1+(n+1)^\lambda} \right]
\end{equation*}
from which
\begin{equation}
\label{S_max_2-A}
S_{N}(\lambda) \to 
\begin{cases}
\widehat{\mathcal{S}}(\lambda)        & \qquad \lambda>1\\
\widehat{A}/N                                 & \qquad \lambda=1\\
\exp\!\left(\!-\frac{N^{1-\lambda}}{1-\lambda}\right)    &\qquad 0<\lambda<1\\
\frac{N}{2^{N-1}}    &\qquad  \lambda=0
\end{cases}
\end{equation}
where $\widehat{\mathcal{S}}(\lambda)=\prod_{n\geq 1}\left[1 -
  \frac{n-1}{n}\,\frac{1}{1+(n+1)^\lambda} \right]$ and the amplitude
$\widehat{A}$ is found by employing the same construction as that used for
Eq.~\eqref{NSN} to yield
\begin{equation*}
\widehat{A} = \prod_{n=1}^\infty\left[1+\frac{2n+1}{n^2(n+2)} \right] 
= \frac{2}{\pi}\cosh\!\left(\frac{\pi\sqrt{3}}{2}\right)
=   4.856379592\ldots
\end{equation*}
Again, the third line in Eq.~\eqref{S_max_2-A} represents only the controlling
factor in the asymptotic behavior in which subdominant contributions as
$N\to\infty$ have been neglected.

The predictions \eqref{S_max_asymp} and \eqref{S_max_2-A} for the two models
\eqref{r} are qualitatively the same for the same values of $\lambda$.  More
precisely, qualitatively different networks emerge only in the extreme case
$\lambda=0$ which is not interesting to us as our goal is to study {\em
  enhanced} redirection. (When $\lambda=0$, the first model in \eqref{r}
leads to uniform attachment without redirection, the process that generates
recursive random trees, while the second model in \eqref{r} leads to constant
redirection probability $r=1/2$ which is equivalent to strictly linear
preferential attachment process.)

\section{Smallest Root Degree}
\label{app:small}

The minimal possible degree of the root is $k=3$ when $N\geq 2$. Using \eqref{root} we get
\begin{equation}
\label{R3:recur}
R(3,N+1)= \left[1-\frac{2-3^{-\lambda}}{N}\right]R(3,N)\,.
\end{equation}
Starting from $R(3,2)=1$ and iterating \eqref{R3:recur} we find
\begin{equation}
\label{R3:sol}
R(3,N)= \frac{\Gamma(N-2+3^{-\lambda})}{\Gamma(N)\,\Gamma(3^{-\lambda})}\simeq 
\frac{1}{\Gamma(3^{-\lambda})}\, N^{-2+3^{-\lambda}}~.
\end{equation}

Let us now probe the behavior of the second smallest degree $k=4$. Using
\eqref{root} we obtain the recurrence
\begin{equation}
\label{R4:recur}
R(4,N+1)= \frac{2-3^{-\lambda}}{N}\,R(3,N)+ \left[1-\frac{3-2^{1-2\lambda}}{N}\right]R(4,N)\,.
\end{equation}
Making the substitution
\begin{equation}
\label{R4:U}
R(4,N) = \frac{\Gamma(N-3+2^{1-2\lambda})}{\Gamma(N)}\,U(N)\,,
\end{equation}
and using \eqref{R3:sol} we recast \eqref{R4:recur} into a simple recursion 
\begin{equation*}
U(N+1) = U(N) + \frac{2-3^{-\lambda}}{\Gamma(3^{-\lambda})}\,\cdot
\frac{\Gamma(N-2+3^{-\lambda})}{\Gamma(N-2+2^{1-2\lambda})}~,
\end{equation*}
from which
\begin{equation}
\label{U:sol}
U(N) = \frac{2-3^{-\lambda}}{\Gamma(2^{1-2\lambda})} + \frac{2-3^{-\lambda}}{\Gamma(3^{-\lambda})}
\sum_{j=1}^{N-3} \frac{\Gamma(j+3^{-\lambda})}{\Gamma(j+2^{1-2\lambda})}~.
\end{equation}
Using Eqs.~\eqref{R4:U} and \eqref{U:sol} one deduces the asymptotic behavior
\begin{equation}
\label{R4:asymp}
R(4,N) \simeq \frac{1}{\Gamma(3^{-\lambda})}\, 
\frac{2-3^{-\lambda}}{1+ 3^{-\lambda}-2^{1-2\lambda}}\,N^{-2+3^{-\lambda}}~.
\end{equation}

The asymptotic behaviors \eqref{R3:sol} and \eqref{R4:asymp} exhibit the same
dependence on the number of nodes $N$. Generally, $R(k,N)\sim
N^{-2+3^{-\lambda}}$ when $N\gg 1$ and $k$ is kept finite.  This, in
conjunction with the scaling form \eqref{scaling}, leads to the small-$x$
behavior given in \eqref{Psi_small} for the scaled root degree distribution.



\newpage

\begin{thebibliography}{99}

\bibitem{Y25} G.~U.~Yule, Phil.\ Trans.\ Roy.\ Soc.\ B {\bf 213}, 21 (1925).

\bibitem{S55} H.~A.~Simon, Biometrika {\bf 42}, 425 (1955).

\bibitem{BA99} A.-L.  Barab\'asi and R. Albert, Science {\bf 286}, 509 (1999).

\bibitem{KRL00} P. L. Krapivsky, S. Redner, and F. Leyvraz, Phys.\ Rev.\
  Lett. {\bf 85}, 4629 (2000).
  
\bibitem{DMS00} S. N. Dorogovtsev, J. F. F. Mendes, and A. N. Samukhin,
  Phys.\ Rev.\ Lett.\ {\bf 85}, 4633 (2000).

\bibitem{DM03} S.~N.~Dorogovtsev and J.~F.~F.~Mendes, {\it Evolution of
    Networks: From Biological Nets to the Internet and WWW} (Oxford
  University Press, Oxford, UK, 2003).

\bibitem{N10} M. E. J. Newman, {\it Networks: An Introduction} (Oxford
  University Press, Oxford, 2010).

\bibitem{FKP02} A. Fabrikant, E. Koutsoupias, and C. H. Papadimitriou,
{\it  Automata, Languages and
Programming},  Lecture Notes in Computer Science,  {\bf 2380}, 110, Springer,
Berlin, 2002).

\bibitem{CBMR04} V. Colizza, J. R. Banavar, A. Maritan, and A. Rinaldo,
  Phys.\ Rev.\ Lett.\ {\bf 92}, 198701 (2004).

\bibitem{B11} M. Barthelemy, Phys.\ Rept.\ {\bf 499}, 1 (2011).

\bibitem{PKSBK12} F. Papadopoulos, M. Kitsak, M. \'Angeles Serrano, M.
  Bogu\~n\'a, and D. Krioukov, Nature {\bf 489}, 537 (2012).

\bibitem{kum} J.~Kleinberg, R.~Kumar, P.~Raghavan, S.~Rajagopalan, and
  A.~Tomkins, in: {\it Proceedings of the International Conference on
    Combinatorics and Computing}, Lecture Notes in Computer Science,
  Vol.~1627, pp.\ 1-18 (Springer-Verlag, Berlin, 1999).

\bibitem{KR01} P. L. Krapivsky and S. Redner, Phys.\ Rev.\ E {\bf 63}, 066123
  (2001).

\bibitem{V03} A. V\'azquez, Phys.\ Rev.\ E {\bf 67}, 056104 (2003).

\bibitem{RA04} H. Rozenfeld and D. ben-Avraham, Phys.\ Rev.\ E {\bf 70},
  056107 (2004).

\bibitem{KR05} P. L. Krapivsky and S. Redner, Phys.\ Rev.\ E {\bf 71}, 036118
  (2005).

\bibitem{LA07} R. Lambiotte and M. Ausloos Europhys.\ Lett. {\bf 77}, 58002
  (2007).

\bibitem{BK10} E. Ben-Naim and P. L. Krapivsky, J. Stat.\ Mech.\ P06004
  (2010).

\bibitem{GR13} A. Gabel and S. Redner, J. Stat.\ Mech.\ P02043 (2013).

\bibitem{KK08} P. L. Krapivsky and D. Krioukov, Phys.\ Rev.\ E {\bf 78},
  026114 (2008).

\bibitem{CS03} R. F. i Cancho and R. V. Sol\'e, \emph{Statistical Mechanics
    of Complex Networks}, no.\ 625 in Lecture Notes in Physics, p.\ 114,
  Springer, Berlin (2003).

\bibitem{BB01} G. Bianconi and A.-L. Bar\'abasi, Phys.\ Rev.\ Lett.  {\bf
    86}, 5632 (2001); G. Bianconi and A.-L. Bar\'abasi, Europhys.\ Lett.\
  {\bf 54}, 436 (2001).

\bibitem{KR02} P. L. Krapivsky and S. Redner, Comput.\ Netw.\ {\bf 39}, 261
  (2002).

\bibitem{BO99} D. L. Bryan and M. E. O'Kelly, J. Regional Science, {\bf 39},
  275 (1999).

\bibitem{H04} J. J. Han, N. Bertain, T. Hao, D. S. Goldberg, G. F. Berriz,
  L. V. Zhang, D. Dupay, A. J. M. Walhout, M. E. Cusick, F. P. Roth, and
  M. Vidal, Nature {\bf 430}, 88 (2004).

\bibitem{GMTA05} R. Guimera, S. Mossa, A. Turtschi, and L. A. N. Amaral,
  Proc.\ Natl.\ Acad.\ Sci.\ USA {\bf 102}, 7794 (2005).

\bibitem{KBM13} J. Kunegis, M. Blattner, and C. Moser, arXiv:1303.6271.

\bibitem{KR02f} P. L. Krapivsky and S. Redner, J. Phys.\ A {\bf 35}, 9517
  (2002).

\bibitem{GKR_13} A. Gabel, P. L. Krapivsky, and S. Redner, Phys.\ Rev.\ E
  {\bf 88}, 050802 (2013).

\bibitem{KR02_b} P. L. Krapivsky and S. Redner, Phys.\ Rev.\ Lett.\ {\bf 89},
  258703 (2002).

\bibitem{G58} E.~J.~Gumbel, {\it Statistics of Extremes} (Columbia University
  Press, New York, 1958).

\bibitem{EP23} F. Eggenberger and G.  P\'olya, Z. Angew.\ Math.\ Mech.\ {\bf
    3}, 279 (1923).

\bibitem{urn} N. Johnson and S. Kotz, \emph{Urn Models and Their
    Applications: An Approach to Modern Discrete Probability Theory}, (Wiley,
  New York, 1977); H. M. Mahmoud, \emph{P\'olya Urn Models} (Chapman \& Hall,
  London, 2008).

\bibitem{DF} B.~Derrida and H.~Flyvbjerg, J.\ Phys.\ A {\bf 20}, 5273 (1987);
  J.\ Physique {\bf 48}, 971 (1987); B.~Derrida and D.~Bessis, J.\ Phys.\ A
  {\bf 21}, L509 (1988).

\bibitem{Higgs} P.~G.~Higgs, Phys.\ Rev.\ E {\bf 51}, 95 (1995).

\bibitem{FIK} L.~Frachebourg, I.~Ispolatov, and P.~L.~Krapivsky, Phys.\ Rev.\
  E {\bf 52}, R5727 (1995).

\bibitem{DJM} B.~Derrida and B.~Jung-Muller, J.\ Stat.\ Phys. {\bf 94}, 277
  (1999).

\bibitem{KGB} P.~L.~Krapivsky, I.~Grosse, and E.~Ben-Naim, Phys.\ Rev.\ E
  {\bf 61}, R993 (2000); P.~L.~Krapivsky, E.~Ben-Naim, and I.~Grosse, J.\
  Phys.\ A {\bf 37}, 2863 (2004).

\bibitem{WS98} D. Watts and S. H. Strogatz, Nature {\bf 393}, 440 (1998).

\bibitem{ABCG06} F. J. Acedo, C. Barroso, C. Casanueva, and J. L. Gal\'an,
  J. Management Studies {\bf 43}, 970 (2006).

\bibitem{FLW08} F. Fu, L. Liu, and L. Wang, Physica A {\bf 387}, 675 (2008).

\bibitem{RB03} E. Ravasz and A.-L. Barab\'asi, Phys.\ Rev.\ E {\bf 67}, 026112 (2003).

\bibitem{AS72} M.~Abramowitz and A.~Stegun, {\it Handbook of Mathematical
    Functions} (Dover, New York, 1965).

\end{thebibliography}
\end{document}